\documentclass[%
 aip,
 amsmath,amssymb,
 reprint,%
]{revtex4-1}

\usepackage{graphicx}
\usepackage{dcolumn}
\usepackage{bm}

\usepackage[utf8]{inputenc}
\usepackage[T1]{fontenc}
\usepackage{mathptmx}
\usepackage{etoolbox}
\usepackage{caption}

\usepackage[frozencache]{minted}
\usepackage{multirow}
\usepackage{dirtree}

\graphicspath{{figures/}}

\usepackage[table]{xcolor}

\makeatletter
\def\@email#1#2{%
 \endgroup
 \patchcmd{\titleblock@produce}
  {\frontmatter@RRAPformat}
  {\frontmatter@RRAPformat{\produce@RRAP{*#1\href{mailto:#2}{#2}}}\frontmatter@RRAPformat}
  {}{}
}%
\makeatother
\begin{document}


\title[SchNetPack 2.0]{SchNetPack 2.0: \\ A neural network toolbox for atomistic machine learning}
\author{Kristof T. Schütt}
\affiliation{Machine Learning Group, Technische Universit\"at Berlin, 10587 Berlin, Germany}
\affiliation{Berlin Institute for the Foundations of Learning and Data, 10587 Berlin, Germany}

\author{Stefaan S. P. Hessmann}
\affiliation{Machine Learning Group, Technische Universit\"at Berlin, 10587 Berlin, Germany}
\affiliation{Berlin Institute for the Foundations of Learning and Data, 10587 Berlin, Germany}

\author{Niklas W. A. Gebauer}
\affiliation{Machine Learning Group, Technische Universit\"at Berlin, 10587 Berlin, Germany}
\affiliation{Berlin Institute for the Foundations of Learning and Data, 10587 Berlin, Germany}
\affiliation{BASLEARN – TU Berlin/BASF Joint Lab for Machine Learning, Technische Universit\"at Berlin, 10587 Berlin, Germany}

\author{Jonas Lederer}
\affiliation{Machine Learning Group, Technische Universit\"at Berlin, 10587 Berlin, Germany}
\affiliation{Berlin Institute for the Foundations of Learning and Data, 10587 Berlin, Germany}

\author{Michael Gastegger}
\email{kristof.schuett@tu-berlin.de, michael.gastegger@tu-berlin.de}
\affiliation{Machine Learning Group, Technische Universit\"at Berlin, 10587 Berlin, Germany}
\affiliation{BASLEARN – TU Berlin/BASF Joint Lab for Machine Learning, Technische Universit\"at Berlin, 10587 Berlin, Germany}

\date{\today}

\begin{abstract}
SchNetPack is a versatile neural networks toolbox that addresses both the requirements of method development and application of atomistic machine learning. 
Version 2.0 comes with an improved data pipeline, modules for equivariant neural networks as well as a PyTorch implementation of molecular dynamics. 
An optional integration with PyTorch Lightning and the Hydra configuration framework powers a flexible command-line interface.
This makes SchNetPack 2.0 easily extendable with custom code and ready for complex training task such as generation of 3d molecular structures.
\end{abstract}

\maketitle

\section{\label{sec:introduction}Introduction}

In recent years, machine learning (ML) techniques have become valuable tools in atomistic modeling of molecules and materials.~\cite{kulik2022roadmap,von2020exploring,keith2021combining,noe2020machine,behler2021four,dral2020quantum}
They have been shown to accurately predict chemical properties~\cite{rupp2012fast,bartok2010gaussian,schuett2017quantum,schutt2017schnet,faber2017prediction,faber2018alchemical,unke2019physnet,huang2020quantum,gasteiger_dimenet_2020} and accelerate molecular dynamics simulations.~\cite{behler2007generalized, behler2015constructing,unke2021machine,batatia2022mace,bartok2018machine}
Machine learning force fields have been applied to systems ranging from small molecules~\cite{schutt2018schnet,chmiela2017machine,chmiela2018towards} over biomolecular systems~\cite{unke2022accurate} to materials with millions of atoms.~\cite{lu202186,musaelian2022learning}
While modeling of potential energy surfaces is arguably the most prominent application, machine learning is integrated into more and more steps of molecular and materials modeling workflows.~\cite{westermayr2021perspective}
It has ventured into the prediction of electronic densities,~\cite{Li2020PRL,Brockherde2017,Fabrizio2019CS} molecular orbitals~\cite{schutt2019unifying, unke2021se} and excited states.~\cite{Ghosh2019AS,Westermayr2020MLST_Perspective,Westermayr2021physically}
Techniques such as reinforcement learning and generative neural networks have enabled complex tasks such as molecular manipulation,~\cite{leinen2020autonomous} or inverse design of 3d molecular structures.~\cite{gebauer2022inverse,kohler2019equivariant,liu2018constrained,simm2021symmetryaware,joshi20213d-scaffold}
Beyond that, unsupervised learning has been applied to learn molecular kinetics~\cite{mardt2018vampnets}, identify chemical moieties~\cite{lederer2022automatic} and learn representations of wavefunctions~\cite{hermann2020deep,pfau2020ferminet}.
These developments come with increasingly diverse demands on an ML toolbox for atomistic modeling.

When the first version of SchNetPack was released~\cite{schuett2018schnetpack}, the aim was to provide a software package that makes neural network potentials easily accessible for researchers in atomistic modeling as well as machine learning.
This has been achieved by making atomistic benchmark sets readily available, establishing a unified structure for neural network potentials and providing a scalable training framework based on PyTorch~\cite{pytorch} that takes large parts of boilerplate code off the researcher.
However, the rapid development of the field described above demands a more flexible approach that is able to adapt to new tasks such as generative models.
Furthermore, the growing PyTorch ecosystem with training frameworks like PyTorch Lightning~\cite{falcon2020pytorchlightning} or Ignite~\cite{pytorch-ignite} makes maintaining an own training framework in SchNetPack an unwarranted technical debt.

\begin{figure*}
\includegraphics[width=.9\textwidth]{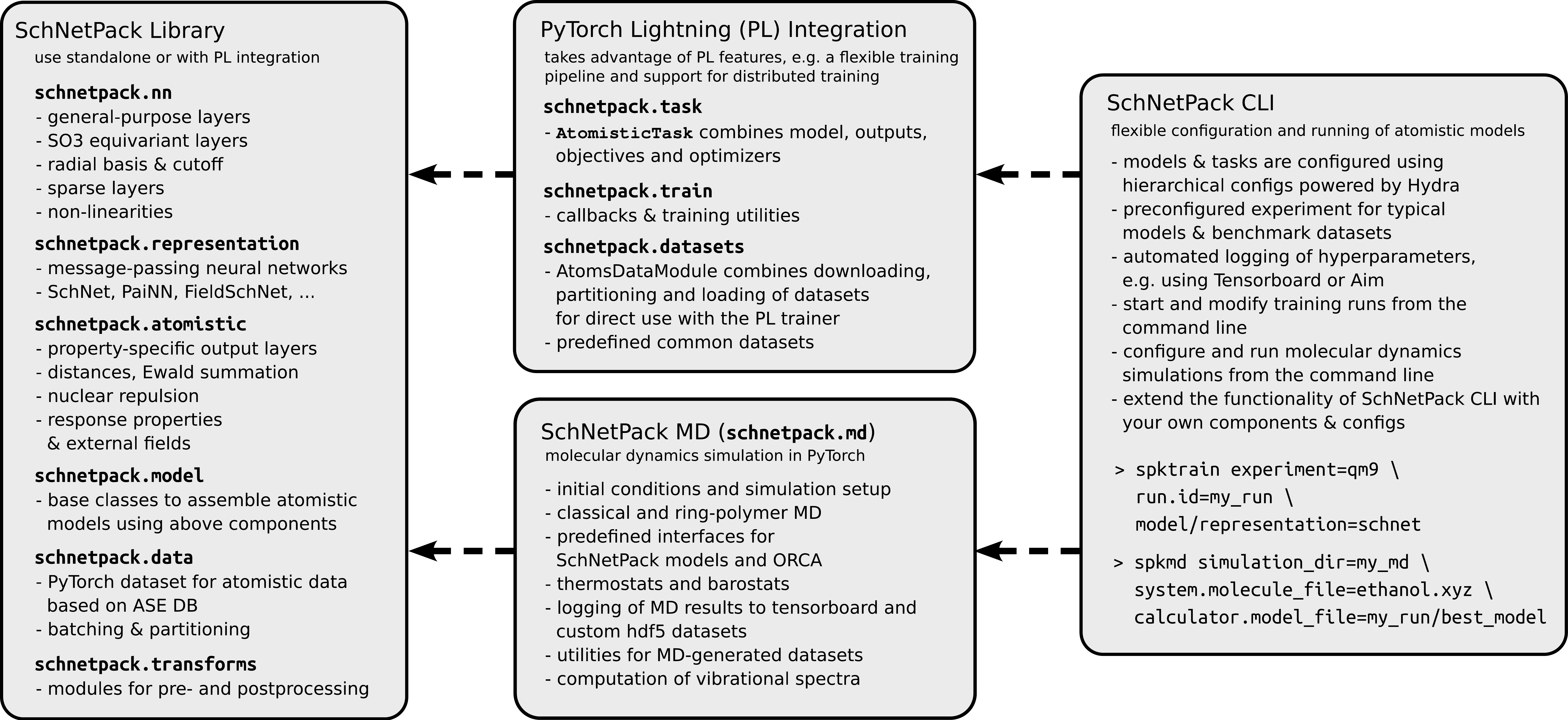}
\caption{\label{fig:overview}Overview of the five major components of the SchNetPack toolbox: the atomistic neural network library, PyTorch Lightning integration, command-line interface and molecular dynamics code. The arrows indicate dependencies between the components, i.e. components can be used independently of components on their right.}
\end{figure*}

With the release of SchNetPack 2.0, major parts of the code base have been rewritten to address the changing demands on a neural network toolbox for atomistic modeling.
Most importantly, SchNetPack is structured into components that can be used individually or as a unified framework.
This makes it straightforward to combine some or all SchNetPack components with other PyTorch-based libraries, such as e3nn~\cite{e3nn_paper} or TorchMD~\cite{doerr2021torchmd}.
Fig.~\ref{fig:overview} gives an overview of the functionalities contained in each component.
The core of the package is the SchNetPack library which consists of the tools required to load atomistic data and define neural network models.
The library includes implementations of representations such as SchNet and PaiNN, but also provides commonly required layers to build custom atomistic representations.
Furthermore, the SchNetPack library includes specific modules for the prediction of common targets, such as energies and response properties.
The library can be used with pure PyTorch or a training framework of choice, as described in detail in Section~\ref{sec:nnlib}.

Alternatively, SchNetPack provides an integration of the library with the PyTorch Lightning (PL) framework~\cite{falcon2020pytorchlightning}.
This enables the use of a plethora of PL features, such as customizable training loops with callbacks, distributed training supporting various accelerators and extensive support for experiment loggers such as Tensorboard~\cite{tensorflow2015-whitepaper} or Aim~\cite{aimui}.
SchNetPack integrates with PL over the \verb+AtomisticTask+, which is composed of models defined using the SchNetPack library, training objectives and optimizers.
We further provide interfaces to popular datasets on the basis of \verb+LightningDataModule+ which enables automatic download and parsing of common benchmark datasets.
The PL interface is described in detail in Section~\ref{sec:plintegration}.

Besides implementing neural networks with the Python API, SchNetPack 2.0 features a command-line interface (CLI) for composing models from the supplied or custom modules, which will be described in Section~\ref{sec:cli}.
It is powered by the Hydra~\cite{Yadan2019Hydra} framework, which allows to build hierarchical YAML configurations.
The structure of this hierarchy is closely oriented on the SchNetPack PL integration, so complex models and training tasks can be described.
Configured training runs can be started and modified from the command line, which makes it easy to quickly scan a large number of hyperparameters and models.
User extensions to SchNetPack can be directly incorporated into the existing CLI.

Finally, SchNetPack 2.0 contains a molecular dynamics code, which makes it possible to directly apply SchNetPack models in simulations with little communication overhead.
Like the rest of the code package, the MD suite is implemented in PyTorch offering full CUDA support.
It retains the batch structure of the neural network toolbox, making it possible to simulate multiple systems in parallel.
Building on this feature, the MD code implements an efficient way to perform ring-polymer MD simulations.
A collection of thermostats and barostats is available to sample different thermodynamic ensembles.
The MD code also features full CLI integration as well as a series of utilities to simplify logging and the analysis of simulation results, e.g. computation of different vibrational spectra.

Code, documentation and tutorials for SchNetPack are available on GitHub\footnote{{https://github.com/atomistic-machine-learning/schnetpack}} and ReadTheDocs~\footnote{{https://schnetpack.readthedocs.io}}.

\section{Neural network library}\label{sec:nnlib}

The SchNetPack 2.0 library provides tools and functionality to build atomistic neural networks and process datasets of molecules and materials.
We have designed the library so that it can be used with vanilla PyTorch, i.e. without the need to integrate with PyTorch Lightning or the Hydra configurations.
Instead we define common interfaces for datasets and models that make them ready to use with the other SchNetPack components.

A major change compared to SchNetPack 1.0 is that the data format is now fully sparse.
Thus, we do no longer have to pad atomic environments with varying number of neighbors.
This required a rewrite of all atomwise operations including the data pipeline as well as the message-passing and output layers.
In the following, we will introduce the improved data pipeline, the pre- and postprocessing modules and the neural network models.

\subsection{Data pipeline}\label{sec:data}
\begin{figure}
	\includegraphics[width=\columnwidth]{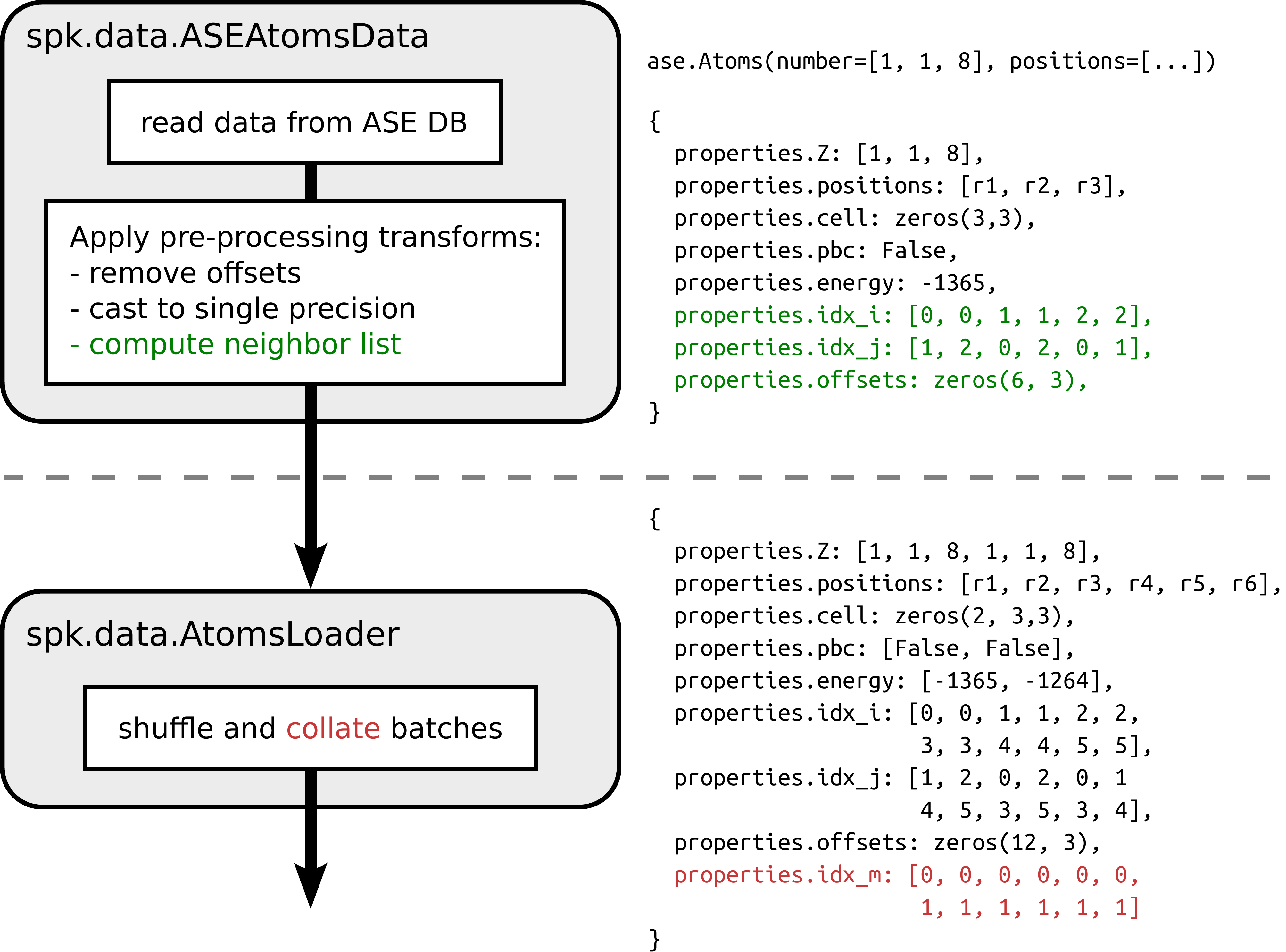}
	\caption{\label{fig:data}The SchNetPack data pipeline. Left: ASEAtomsData provides an interface to data stored in an ASE DB and applies a sequence of preprocessing transforms. The AtomsLoader loads this data with multiprocessing and builds batches to be passed to the model. Right: example of the input dictionaries after preprocessing (top) and batching (bottom). All values of the dictionary are PyTorch tensors.}
\end{figure}

The SchNetPack data pipeline mainly consists of \verb+ASEAtomsData+, adhering to the PyTorch dataset inferface, and \verb+AtomsLoader+, being a PyTorch data loader with a customized collate function for batching of atomistic data.
An important feature of \verb+ASEAtomsData+ is that one can provide a list of preprocessing transforms.
They are applied in sequence to single data instances, before the latter are batched in the \verb+AtomsLoader+.
This is particularly useful for calculating neighbor lists, removing offsets or casting properties.
The transforms will be described in detail in Section~\ref{sec:preprocessing}.
Fig.~\ref{fig:data} gives an overview of the data pipeline and the format of the data before and after batching.

As in SchNetPack 1.0, the default PyTorch dataset \verb+ASEAtomsData+ is based on the database of the Atomic Simulation Environment (ASE).~\cite{ase}
However, the internal format has slightly changed in that it expects a dictionary of units for all properties and the atom positions. 
This enables automatic unit conversion without requiring internal default units.
We have included the script \verb+spkconvert+ to add this information to older datasets.
The following code snippet demonstrates how to create a new dataset:
\begin{small}
\begin{minted}[python3=true, xleftmargin=0pt, numbersep=0pt, numbers=none, mathescape]{python}
# create new dataset
new_dataset = ASEAtomsData.create(
 './new_dataset.db', distance_unit='Ang',
 property_unit_dict={
   'energy':'kcal/mol', 'forces':'kcal/mol/Ang'
 },
 atomrefs = {
   'energy': [0.0, -313.5150902000774, ...]
 }
)

atoms_list = [
  ase.Atoms(numbers=[1, 1, 8], positions=[...]), ...
]
property_list = [{
  'energy': np.array([-450.2]),
  'forces': np.array([[0.8, -0.4, 1.3], ...])}
]
new_dataset.add_systems(property_list, atoms_list)
# add metadata for custum train/test split
new_dataset.update_metadata(
  train_idx=[0, 12, 53, ...])

# create training dataset and retrieve an entry
train_data:ASEAtomsData = new_dataset.subset(
  new_dataset.metadata["train_idx"]
)
some_molecule = train_data[2]
\end{minted}
\end{small}
The atomistic systems are added by providing lists of ASE \verb+Atoms+ and dictionaries of NumPy~\cite{numpy} arrays.
Additionally, we may provide single-atom reference energies (\verb+atomrefs+), which can be used for preprocessing the target properties.
Both the single-atom energies and property units are stored as metadata in the ASE DB.
Beyond that, additional custom metadata can be stored, such as predefined train / test splits in the code example above.

\verb+ASEAtomsData+ is an implementation of the abstract base class \verb+BaseAtomsData+, which defines a general interface to SchNetPack datasets.
This makes it possible to extend SchNetPack with custom data formats, for example for distributed datasets or special data types such as wave function files.
Independent of the concrete implementation of \verb+BaseAtomsData+, the format of retrieved data is a dictionary mapping from strings to PyTorch tensors, as shown in the example in Fig.~\ref{fig:data} (right).

The \verb+AtomsLoader+ batches the preprocessed inputs after optional shuffling.
Since systems can have a varying number of atoms, the batch dimension for atom-wise properties, such as positions and forces, runs over atoms instead of systems.
The index of the corresponding system in the batch is encoded in the PyTorch tensor \verb+idx_m+.
Index tensors, e.g. generated by neighbor lists (\verb+idx_i+, \verb+idx_j+), have to be treated differently since they refer to the atom indices within a single data example.
Therefore, the collate function shifts the indices to refer to the correct position within the batch.

\subsection{Pre- and postprocessing transforms}\label{sec:preprocessing}

\begin{table*}
\renewcommand\arraystretch{1.2}
\caption{\label{tab:transforms} List of pre- and postprocessing transforms.}
\begin{ruledtabular}
\begin{tabular}{lllp{0.55\linewidth}}
\textbf{\textit{Category}} & \textbf{\textit{Transform}} & \textbf{\textit{Usage}} & \textbf{\textit{Description}}\\
\hline
\textbf{Neighbor lists} & MatScipyNeighborList & Pre & Neighbor list implementation based on Matscipy~\cite{matscipy}. This should be preferred. \\
 & ASENeighborList & Pre & Neighbor list based on Atomic Simulation Environment.\\
  & TorchNeighborList & Pre & Neighbor list implemented in PyTorch. \\
  & CachedNeighorList & Pre & Wrapper for other neighbor list transforms that caches the results. \\
  & SkinNeighborList & Pre & Wrapper around neighbor list transform that only recalculates neighbor indices after atom positions change more than a given threshold. This can be useful for structure relaxation. \\
  & FilterNeighbors & Pre & Filter previously calculated neighbor indices. \\
  & CountNeighbors & Pre & Count \& store number of neighbors for each atom. \\
  & WrapPositions & Pre & Wrap atom position into periodic cell. \\ \hline
\textbf{Casting} & CastMap & Pre / Post & Cast all properties according to supplied type map. \\
& CastTo32 & Pre / Post & Cast all double precision inputs to single precision \\
& CastTo64 & Pre / Post & Cast all single precision inputs to double precision \\ \hline
\textbf{Scale \& Offset} & ScaleProperty & Pre / Post & Scale an input or result by data mean, standard deviation or given factor \\
& RemoveOffsets & Pre / Post & Remove single-atom reference and/or mean from an input or result \\
& AddOffsets & Pre / Post & Add single-atom reference and/or mean to an input or result \\
\end{tabular}
\end{ruledtabular}
\end{table*}

SchNetPack transforms are PyTorch modules that have no trainable parameters and are used for preprocessing of inputs or postprocessing of model results.
Preprocessing transforms are applied before batching, i.e. they operate on single inputs.
For example, virtually every SchNetPack model requires a preprocessing transform that constructs a neighbor list.
However, different types of postprocessing may be demanded in the training and prediction phases.
For example, a preprocessor might need to perform data augmentation during training, but not during predicting.
Another example is the \verb+SkinNeighboList+ which takes advantage of structural similarity of sequential examples, which can be encountered in prediction tasks such as molecular dynamics simulations or structure relaxation, but not during training.

Postprocessing transforms act on batches in the result dictionary and are part of the \verb+AtomisticModel+ described in Section~\ref{sec:model}.
However, the loss function is supposed to be independent of postprocessing.
Thus, these transforms are only enabled for prediction but not during training and evaluation.
The currently supported pre- and postprocessing transforms are listed in Table~\ref{tab:transforms}.

In the following, we illustrate the usage of transforms at the use case of casting between single and double precision: On the one hand, double precision is required to accurately represent the comparatively small energy differences compared to the much larger scale of the total energy. On the other hand, single or even half precision is needed for fast processing of neural networks on GPUs. We solve this by applying a chain of pre- and postprocessing transforms.
First, offsets such as single-atom energies are removed from the energy targets using the \verb+RemoveOffsets+ transform at double precision.
Only then do we use the \verb+CastTo32+ transform to cast all floating point inputs to single precision.
Thus, the neural network is trained to predict the target without these offsets.
The preprocessing modules can be set manually as follows:
\begin{small}
\begin{minted}[python3=true, xleftmargin=0pt, numbersep=0pt, numbers=none, mathescape]{python}
atomrefs = train_data.atomrefs
train_data.transforms = [
  RemoveOffsets(
    remove_atomrefs=True, 
    atomrefs=train_data.atomrefs
  ),
  MatScipyNeighborList(cutoff=5.),
  CastTo32(),
]
\end{minted}
\end{small}
Alternatively, obtaining single-atom references or other data-dependent initialization can be taken care of automatically when using PyTorch Lighting, as described in Section~\ref{sec:plintegration}.

For the final predictions, the postprocessing modules are activated, first casting the neural network prediction to double precision (\verb+CastTo64+) and adding the single-atom energies afterwards (\verb+AddOffsets+).

\subsection{Atomistic models}\label{sec:model}

\begin{figure}
\includegraphics[width=\columnwidth]{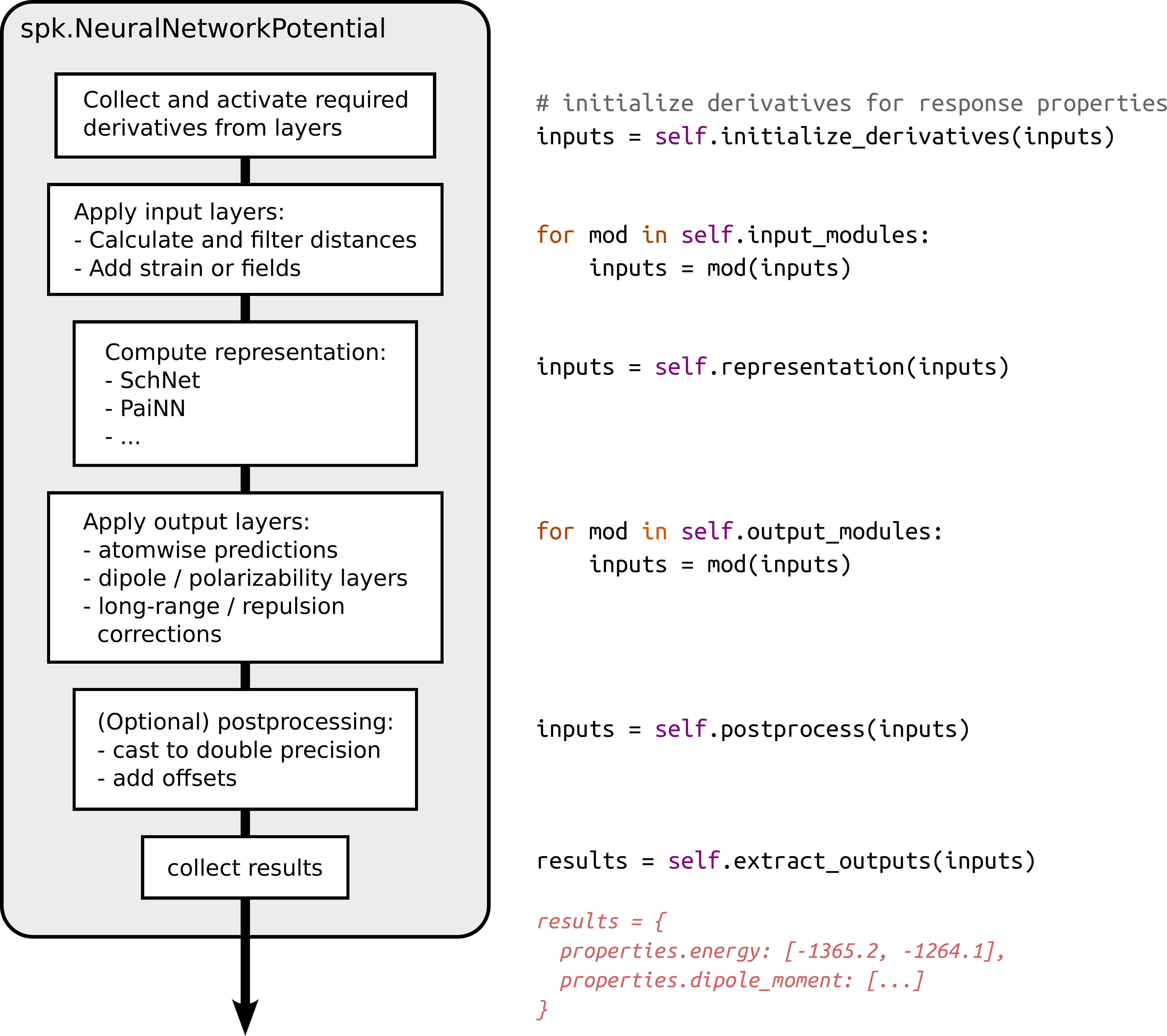}
\caption{\label{fig:nnp}Data processing in \texttt{NeuralNetworkPotential} with corresponding code.}
\end{figure}

All neural networks implemented with SchNetPack are supposed to subclass \verb+AtomisticModel+.
It is a PyTorch module with additional functionality that is commonly required for atomistic machine learning.
In particular, it offers support for the previously described postprocessors, filtering of result dictionaries as well as a convenient mechanism to initialize and collect automatic derivatives.

While \verb+AtomisticModel+ can be used to implement neural networks for a broad class of tasks, SchNetPack also includes its more structured \verb+NeuralNetworkPotential+.
This is tailored towards ML potentials and makes use of all the features of \verb+AtomisticModel+.
We recommend using \verb+NeuralNetworkPotential+ whenever possible as it allows for an easier integration with the SchNetPack CLI.
On the other hand, \verb+AtomisticModel+ may be employed for more general tasks.
This will be demonstrated at the example of the generative model cG-SchNet in Section~\ref{sec:cgschnet}.

Fig.~\ref{fig:nnp} depicts the processing flow of \verb+NeuralNetworkPotential+.
The main concept is to pass along an inputs dictionary that is modified by submodules.
This enables the definition of complex neural networks with multiple inputs and outputs as just a sequence of modules.
An immediate advantage is that such a model can be easily composed using configuration files and the SchNetPack CLI.

Many neural network potentials are employed for predicting response variables such as atomic forces or electric multipoles.
Therefore, it is often required to calculate derivatives of model outputs w.r.t. input variables.
The derivative tracking of these input variables needs to be activated before any of the modules are applied.
The method \verb+AtomisticModel.collect_derivatives()+ can be called during initialization to scan all submodules of \verb+NeuralNetworkPotential+ for required derivatives.
Based on this list, we enable the generation of corresponding backwards graphs using the \verb+AtomisticModel.initialize_derivatives()+ method during the forward pass.

In the next step, the dictionary is passed to a sequence of so-called \verb+input_modules+.
These may carry out preparatory steps such as calculating distances, applying strain to a lattice or adding auxiliary inputs such as external fields.
The inputs then are passed on to the \emph{representation} module, which constructs atom-wise features based on the input dictionary. 
This is often achieved by a message passing neural network that is designed to be equivariant to the symmetries of an atomistic structure (see Section~\ref{sec:representation}).
The resulting features can be scalars, vectors or general multipoles and are stored under a corresponding key in the dictionary.

Finally, a sequence of \verb+output_modules+ computes the final results. For example, energies are often predicted as a sum over atom-wise energy contributions, which is achieved by the \verb+Atomwise+ module. We refer to both input and output modules as \emph{atomistic modules}, since they are usually tailored to the specifics of atomistic data -- in contrast to general purpose layers such as convolutions.
We describe the supported operations in detail in Section~\ref{sec:outputmods}.

Before the results are returned, \verb+NeuralNetworkPotential+ makes use of two more features of \verb+AtomisticModel+:
First, a sequence of postprocessors is applied, if they have been enabled. This is usually only done in prediction mode, but not during training.
Second, the input dictionary was updated without removing information as it passed through the model.
Therefore, it still contains all the raw inputs and intermediate features.
The \verb+extract_outputs+ method filters only the results that are supposed to be returned by the model.
This is achieved in an semi-automatic fashion by scanning all submodules for potential model outputs during initialization.

\subsection{Message passing and equivariant neural networks}\label{sec:representation}

\begin{table*}
\renewcommand\arraystretch{1.7}
\caption{\label{tab:reprmods} List of modules for message passing and equivariant neural networks.}
\begin{ruledtabular}
\begin{tabular}{llp{0.55\linewidth}}
\textbf{\textit{Category}} & \textbf{\textit{Module}} & \textbf{\textit{Description}}\\
\hline
\textbf{Radial basis} & \verb+GaussianRBF+~\cite{behler2011atom,schuett2017quantum} & Gaussian radial basis functions centered on an equidistant grid in the interval [start, cutoff]\\
& \verb+GaussianRBFCentered+~\cite{behler2011atom} & Gaussian radial basis functions centered at zero with varying widths \\
 & \verb+BesselRBF+~\cite{gasteiger_dimenet_2020} & $0^\text{th}$-order Bessel radial basis functions: $f(r) = \sin(\omega r) / r$ \\ \hline
\textbf{Cutoff} & \verb+CosineCutoff+~\cite{behler2011atom} & 
$
f(r) = \tfrac{1}{2} \left[1 + \cos\left(\frac{\pi r}{r_\text{cutoff}}\right)\right], \quad
           r < r_\text{cutoff}
$ \\ 
& \verb+MollifierCutoff+ & 
$
f(r) = \exp\left[1 - \frac{1}{1 - \left(\frac{r}{r_\text{cutoff}}\right)^2} \right], \quad
           r < r_\text{cutoff}
$ \\ \hline
\textbf{Nonlinearity} & \verb+shifted_softplus+~\cite{schutt2017schnet} & $f(x) = \ln(0.5 + 0.5 e^{-x})$ \\ \hline
\textbf{Sparse} & \verb+scatter_add+ & Sum over a sparse dimension, e.g. for sums of atoms and neighbors as well as Clebsch-Gordon tensor products \\ \hline
\textbf{SO(3) equivariance} & \verb+RealSphericalHarmonics+ & Generates the real spherical harmonics for a batch of unit vectors \\
& \verb+SO3TensorProduct+\cite{thomas2018tensor} & $f(x, y)_{(l m)} = \sum\limits_{l_1 m_1} \sum\limits_{l_2 m_2} x_{l_1 m_1} y_{l_2 m_2} C_{l_1 m_1 l_2 m_2}^{l m}$ \\
& \verb+SO3Convolution+\cite{thomas2018tensor} & $f(x)_{i(l m)} = \sum\limits_{j \in nbh[i]} \sum\limits_{l_1 m_1} \sum\limits_{l_2 m_2} x_{j(l_1 m_1)} W_{l_2}(r_{ij}) Y_{l_2,m_2}(\vec{r}_{ij}/r_{ij}) C_{l_1 m_1 l_2 m_2}^{l m}$ \\
& \verb+SO3GatedNonlinearity+ & $f(x)_{i(l m)} = x_{i(l, m)} \sigma (x_{i(0, 0)})$ \\
& \verb+SO3ParametricGatedNonlinearity+\cite{weiler20183d} & $f(x)_{i(l m)} = x_{i(l, m)} \sigma (g_\theta(x_{i(0, 0)}))$ \\
\end{tabular}
\end{ruledtabular}
\end{table*}

SchNetPack provides the tools to build a wide variety of atomistic machine learning models, even beyond neural networks.
However, our focus remains on end-to-end neural networks that build atom-wise representations.
In recent years, the two concepts that have dominated this field are neural message passing~\cite{schuett2017quantum,gilmer2017neural} and equivariant neural networks~\cite{thomas2018tensor,weiler20183d}.
SchNetPack comes with four implemented atomistic representations: \begin{itemize}
    \item \textbf{SchNet}\cite{schutt2017schnet,schutt2018schnet}: The name-giving continuous-filter convolutional network motivated the creation of SchNetPack. It uses rotationally-invariant filters and, although it is no longer the most accurate model five years after it was first proposed, it is pretty lightweight.
    \item \textbf{FieldSchNet}\cite{gastegger2021machine}: An extension of SchNet that makes use of atomic dipole features to handle external fields in order to model response properties and solvent effects.
    \item \textbf{PaiNN}\cite{schutt2021equivariant}: The successor to SchNet that uses equivariant message passing in Cartesian space.
    \item \textbf{SO3net}: A minimalist SO(3)-equivariant neural network in the spirit of Tensor Field Networks~\cite{thomas2018tensor} or NequIP~\cite{batzner2022nequip} that showcases the spherical harmonics and Clebsch-Gordan tensor product modules of SchNetPack.
\end{itemize}

Table~\ref{tab:reprmods} gives an overview of related modules that we provide for building atomistic representations.
This includes radial basis and cutoff functions, nonlinearities and SO(3) equivariant layers.
Naturally, these are only additions to the large variety of modules already included in PyTorch, e.g. the SiLU nonlinearity~\cite{hendrycks2016gaussian}, which has been employed in many recent atomistic neural networks~\cite{gasteiger_dimenet_2020,schutt2021equivariant,batzner2022nequip}.

By convention, the atomistic representations in SchNetPack have the shape $(n_\text{atoms}, [n_\text{spatial},] n_\text{features})$, where the first dimension runs over all atoms in the batch, the second is reserved for optional directional channels of equivariant representations and the last is reserved for feature channels.
In this format, it is straightforward to define message passing operations using the atom and neighbor indices obtained from neighbor list transforms during preprocessing. For example, the continuous-filter convolution of SchNet~\cite{schutt2017schnet} can be implemented as follows:
\begin{small}
\begin{minted}[python3=true, xleftmargin=0pt, numbersep=0pt, numbers=none, mathescape]{python}
# inputs: 
# x: atom-wise representation
# Wij: radial filters
# idx_i, idx_j: neighbor list indices 
x_j = x[idx_j]
x_ij = x_j * Wij
y = scatter_add(x_ij, idx_i, dim_size=x.shape[0])
\end{minted}
\end{small}

The second (optional) dimension of atom-wise representations is reserved for the directional features of equivariant neural networks.
For the vectorial features, e.g. in case of PaiNN, this corresponds to the Cartesian $(x,y,z)$-directions ($n_\text{spatial} = 3$).
In case of SO(3)-equivariant features, the dimensions correspond to the real spherical harmonics $Y_{l,m}$ as follows: $[Y_{0, 0}, Y_{1,-1}, Y_{1,0}, Y_{1,1}, Y_{2,-2}, Y_{2,-1}, \dots]$.

For the implementation of tensor product modules required for SO(3)-equivariant models, the Clebsch-Gordan coefficients $C_{l_1 m_1 l_2 m_2}^{l m}$ are precomputed during initialization and stored in sparse format with the non-zero coefficients \verb+clebsch_gordan+ and three combined index tensors \verb+idx_in_1+, \verb+idx_in_2+ and \verb+idx_out+ corresponding to the tuples $(l_1,m_1)$, $(l_2, m_2)$ and $(l,m)$, respectively.
This enables fast calculation of tensor products. For example, the \verb+SO3TensorProduct+ module calculating
\[
f(x, y)_{(l m)} = \sum\limits_{l_1 m_1} \sum\limits_{l_2 m_2} x_{l_1 m_1} y_{l_2 m_2} C_{l_1 m_1 l_2 m_2}^{l m},
\]
can be implemented as follows:
\begin{small}
\begin{minted}[python3=true, xleftmargin=0pt, numbersep=0pt, numbers=none, mathescape]{python}
# x, y: atom-wise features $[n_\text{atoms}, (l_\text{max}+1)^2, n_\text{features}]$
h = x[:, idx_in_1, :] \
    * y[:, idx_in_2, :] \
    * clebsch_gordan[None, :, None]
z = scatter_add(h, idx_out, 
        dim_size=(lmax + 1) ** 2, dim=1)
\end{minted}
\end{small}
The combined indices \verb+idx_in_1+ and  \verb+idx_in_2+ select the spherical basis functions to be multiplied by the non-zero Clebsch-Gordan coefficients.
The third index tensor is used for summation of terms mapping to the same $(l,m)$ using \verb+scatter_add+ analogous to the atom accumulation in the message passing shown above.
Similarly, \verb+SO3Convolutions+ or other custom SO(3)-equivariant modules can be implemented in a straightforward fashion.

\subsection{Atomistic modules}\label{sec:outputmods}

\begin{table*}
\renewcommand\arraystretch{1.7}
\caption{\label{tab:atomistic} List of atomistic modules for preparing the raw inputs and predicting various properties.}
\begin{ruledtabular}
\begin{tabular}{llp{0.6\linewidth}}
\textbf{\textit{Category}} & \textbf{\textit{Module}} & \textbf{\textit{Description}}\\
\hline
\textbf{Distances} & \verb+PairwiseDistances+ & Compute pair-wise distances from indices provided by a neighbor list transform. \\
& \verb+FilterShortRange+ & Separate distances below a short-range cutoff from all supplied distances. \\ \hline
\textbf{Properties} & \verb+Atomwise+~\cite{behler2007generalized} & Predicts a property from atom-wise contributions and accumulates global prediction using sum or average, e.g. for the energy $E = \sum\limits_{i=1}^{n_\text{atoms}} E(x_i)$ \\
& \verb+DipoleMoment+\cite{gastegger2017machine,veit2020predicting,schutt2021equivariant} & Predicts dipole moments from latent partial charges and (optionally) local, atomic dipoles:
\[
\vec{\mu} = \sum\limits_{i=1}^{n_\text{atoms}} q(x_i) \vec{r}_i \left[ + \vec{\mu}_\text{atomic}(\vec{x}_i) \right]
\] \\
& \verb+Polarizability+~\cite{schutt2021equivariant} & Predicts polarizability using the tensor rank factorization:
\[
\alpha = \sum_{i=1}^N \alpha_\text{0}(x_i) \, I_3 +\vec{\nu}(\vec{x}_i) \otimes \vec{r_i} + \vec{r_i} \otimes \vec{\nu}(\vec{x}_i)
\]
\\
& \verb+Aggregation+ & Aggregate predictions from multiple atomistic modules to a single output variable, e.g. calculating the energy prediction as the sum of short-range and long-range energies.
\\ \hline
\textbf{Potentials} & \verb+EnergyCoulomb+ & Compute Coulomb energy from a set of (latent) point charges.
\\
& \verb+EnergyEwald+ & Compute the Coulomb energy of a set of (latent) point charges inside a periodic box using Ewald summation.
\\
& \verb+ZBLRepulsionEnergy+~\cite{ziegler1985stopping,unke2021spookynet} & Computes a Ziegler-Biersack-Littmark-style repulsion energy.
\\ \hline
\textbf{Response} & \verb+StaticExternalFields+ & Input module for setting up dummy external fields. These do not receive inputs, but are only used to calculate response properties with autograd.
\\
& \verb+Strain+ & Input module for setting up dummy strain. It does not receive inputs, but is only used to calculate the stress tensor using autograd.
\\
& \verb+Forces+ & Output module that predicts forces and stress as response of the energy prediction w.r.t. the atom positions and strain.
\\
& \verb+Response+~\cite{gastegger2021machine} & Output module that computes different response properties by taking derivatives of an energy model. Supports forces, stress, Hessian, dipole moment (and its derivatives) polarizability (and its derivatives), shielding tensor as well as nuclear spin-spin coupling.
\end{tabular}
\end{ruledtabular}
\end{table*}

\emph{Atomistic modules} are specific to the properties to be predicted and informed by the regularities of the underlying physics.
In contrast to general purpose layers, such as non-linearities or convolutions, they are supposed to directly operate on the input dictionary which is passed through the model, as described in Section~\ref{sec:model}.
This makes it easy to compose complex atomistic neural networks.
Thus, the input and output modules of a \verb+NeuralNetworkPotential+ are usually atomistic layers.

Table~\ref{tab:atomistic} describes the atomistic modules currently supported by SchNetPack.
A range of output modules is concerned with predicting atomistic properties from the previously generated representations.
The most common is \verb+Atomwise+, which predicts a property as a sum of atom-wise contributions, e.g. when predicting potential energy surfaces.~\cite{behler2007generalized,bartok2010gaussian,schuett2017quantum}
We have also included specialized layers for tensorial properties such as dipole moments and polarizability tensors.
Beyond that, it is often helpful for generalization to include physically-inspired terms in the energy prediction.
This is includes electrostatic modules and ZBL potentials~\cite{ziegler1985stopping} for nuclear-nuclear repulsion.

Virtually every neural network potential requires the \verb+PairwiseDistances+ module as an input layer, which makes use of the indices calculated by a neighbor list during preprocessing.
The distance calculation being part of the model enables straightforward automatic differentiation w.r.t. to atom positions.
The \verb+Forces+ module provides atomic forces and the stress tensor as derivatives w.r.t. atomic positions and strain.
Beyond that, SchNetPack includes the \verb+Response+ module, which additionally supports response properties w.r.t. external (electric or magnetic) fields as well as higher-order derivatives, e.g. for polarizability or shielding tensors.
The FieldSchNet example in Section~\ref{sec:fieldschnet} demonstrates how this can be employed in practice.

\section{PyTorch Lightning integration}\label{sec:plintegration}

While it is possible to use the previously described SchNetPack library on its own, a third-party framework that takes care of the boilerplate code required for training and validation can significantly speed up the development process.
We chose PyTorch Lightning~\cite{falcon2020pytorchlightning} as the default training framework for SchNetPack, as it supports a wide variety of hardware devices and distribution strategies.
In the following, we describe how \verb+LightningModule+ and \verb+LightningDataModule+ are employed to predefine common datasets, tasks and workflows.


The PyTorch Lightning trainer expects a \verb+LightningModule+ which defines the learning task, i.e. a combination of model definition, objectives and optimizers.
SchNetPack provides the \verb+AtomisticTask+, that integrates the \verb+AtomisticModel+, as described in Section~\ref{sec:model}, with PyTorch Lightning.
The task configures the optimizer, defines the training, validation and test steps, calculates the loss and logs training metrics.
For this purpose, \verb+AtomisticTask+ expects a list of of \verb+ModelOutput+s, which map a target property to a key in the results dictionary of the \verb+AtomisticModel+.
Additionally, \verb+ModelOutput+ requires a loss function and (optionally) a list of \verb+Metric+s that are tracked during training and validation.
Given a model \verb+my_neural_network_potential+ that stores the predictions \verb+energy+ and \verb+forces+ in the result dictionary, the task of training a neural network potential can be defined as follows:
\begin{small}
\begin{minted}[python3=true, xleftmargin=0pt, numbersep=0pt, numbers=none, mathescape]{python}
output_energy = ModelOutput(
    name="energy",
    loss_fn=torch.nn.MSELoss(),
    loss_weight=0.01,
    metrics={"MAE": torchmetrics.MeanAbsoluteError()}
)

output_forces = ModelOutput(
    name="forces",
    loss_fn=torch.nn.MSELoss(),
    loss_weight=0.99,
    metrics={"MAE": torchmetrics.MeanAbsoluteError()}
)
task = AtomisticTask(
    model=my_neural_network_potential,
    outputs=[output_energy, output_forces],
    optimizer_cls=torch.optim.AdamW,
    optimizer_args={"lr": 1e-4}
)
\end{minted}
\end{small}

To define a dataset, we subclass PyTorch Lightning data modules with \verb+AtomsDataModule+. 
This combines the data classes introduced in Section~\ref{sec:data} with code for preparation, setup and partitioning into train, validation and test splits.
The default splitting strategy is random sampling, however other strategies, such as sub-sampling predefined partitions or keeping certain groups of structures in the same partition, are supported as well.
One may provide separate preprocessing transforms for train, validation and test splits if needed, e.g. when data augmentation is only required for the training data.

Beyond that, data modules may take care of setting up the data for distributed training, e.g. copying the data to a local storage.
We provide specialized \verb+AtomsDataModule+s for common benchmark sets, that automatically download and parse the data.
Currently supported datasets include \verb+QM9+~\cite{qm9two}, \verb+(r)MD17+~\cite{chmiela2017machine,christensen2020role}, \verb+MD22+~\cite{chmiela2022accurate}, \verb+OMDB+~\cite{olsthoorn2019band}, and the \verb+MaterialsProject+~\cite{mp1}.
Additional datasets can be added by sub-classing \verb+AtomsDataModule+ and overriding the \verb+prepare_data+ method with custom code for downloading and parsing.

Here is an example of how the \verb+MD17+ datamodule can be used:
\begin{small}
\begin{minted}[python3=true, xleftmargin=0pt, numbersep=0pt, numbers=none, mathescape]{python}
ethanol_data = MD17(
    "/path/to/data.db",
    molecule='ethanol',
    batch_size=10,
    num_train=1000,
    num_val=1000,
    transforms=[
        MatScipyNeighborList(cutoff=5.),
        RemoveOffsets(MD17.energy, remove_mean=True),
        CastTo32()
    ],
    num_workers=1
)
ethanol_data.prepare_data()
ethanol_data.setup()

# iterate over training batches
for batch in ethanol_data.train_dataloader():
    print(batch)
\end{minted}
\end{small}
The passed transforms are applied to all partitions in this case and \verb+RemoveOffsets+ is automatically initialized with the training data statistics.
Manual calling of \verb+prepare_data+, which downloads and parses the data, and \verb+setup+, which creates and loads the partitions, is necessary here because we retrieve the data loader and iterate over the training data.
Instead, one may pass the data module directly to the PyTorch Lightning trainer class, which ensures that \verb+prepare_data+ is called exactly once.
However, the \verb+setup+ method is called in every process of distributed training.

Finally, we put everything together by passing the task and data module to the PyTorch Lightning trainer, which executes the training loop:
\begin{small}
\begin{minted}[python3=true, xleftmargin=0pt, numbersep=0pt, numbers=none, mathescape]{python}
trainer = pl.Trainer()
trainer.fit(task, datamodule=ethanol_data)
\end{minted}
\end{small}
The training process can be adapted by callbacks and loggers as well as specifying options to train on several (distributed) devices.
Please refer to the PyTorch Lightning documentation~\footnote{https://pytorch-lightning.readthedocs.io} for more information.

\section{Configuration and command-line interface}\label{sec:cli}
\begin{figure}[tb]
\begin{scriptsize}
\begin{minted}[xleftmargin=10pt, numbersep=5pt, mathescape, linenos]{yaml}
defaults:
  - override /model: nnp
  - override /data: qm9

run:
  experiment: qm9_${globals.property}

globals:
  cutoff: 5.
  lr: 5e-4
  property: energy_U0
  aggregation: sum

data:
  transforms:
    - _target_: schnetpack.transform.SubtractCenterOfMass
    - _target_: schnetpack.transform.RemoveOffsets
      property: ${globals.property}
      remove_atomrefs: True
      remove_mean: True
    - _target_: schnetpack.transform.MatScipyNeighborList
      cutoff: ${globals.cutoff}
    - _target_: schnetpack.transform.CastTo32

model:
  output_modules:
    - _target_: schnetpack.atomistic.Atomwise
      output_key: ${globals.property}
      n_in: ${model.representation.n_atom_basis}
      aggregation_mode: ${globals.aggregation}
  postprocessors:
    - _target_: schnetpack.transform.CastTo64
    - _target_: schnetpack.transform.AddOffsets
      property: ${globals.property}
      add_mean: True
      add_atomrefs: True

task:
  outputs:
    - _target_: schnetpack.task.ModelOutput
      name: ${globals.property}
      loss_fn:
        _target_: torch.nn.MSELoss
      metrics:
        mae:
          _target_: torchmetrics.regression.MeanAbsoluteError
        rmse:
          _target_: torchmetrics.regression.MeanSquaredError
          squared: False
      loss_weight: 1.
\end{minted}
\end{scriptsize}
\caption{SchNetPack experiment config for prediction tasks of QM9 properties with the \texttt{Atomwise} output layer.\label{fig:qm9config}}
\end{figure}

SchNetPack training runs can be defined using the hierarchical configurations framework Hydra~\cite{Yadan2019Hydra}.
This enables configuration of complex neural network potentials using YAML files, provides powerful command-line tools and makes it easy for developers to extend SchNetPack with external code.

The main command of the SchNetPack CLI is \verb+spktrain+, which creates a new run directory and starts the training of a configured model.
We refer to a predefined configuration including model, task, data, etc. as an \emph{experiment}, which can be started as follows:
\begin{small}
\begin{minted}[xleftmargin=0pt, numbersep=0pt, numbers=none, mathescape]{bash}
spktrain experiment=qm9_atomwise
\end{minted}
\end{small}
This starts the training with the default settings for energies of the QM9 dataset.
The script first prints the flattened config of the run, i.e. the config when specified in a single YAML file.
Fig.~\ref{fig:qm9config} instead shows the hierarchical experiment configuration that this has been derived from.

\subsection{Structure of the configuration}
The structure of configurations is heavily oriented on the building blocks introduced in Sections~\ref{sec:nnlib} and \ref{sec:plintegration}.
Due to its hierarchy, we only have to override the defaults and make use of reusable, predefined config groups, e.g. for the model or the dataset.
The major config groups of SchNetPack are:
\begin{itemize}
    \item \verb+run+: Definition of run-specific variables, such as the run \verb+id+ as well as directories, where the metrics are logged and the trained model and the data will be stored. If no run id is given, SchNetPack will create a unique hash.
    \item \verb+globals+: Custom variables to be used across the whole config can be added here. This is possible through the use of the interpolation syntax \verb+${globals.variable}+.
    \item \verb+data+: Definition of the dataset as described by \verb+AtomsDataModule+ (see Section~\ref{sec:plintegration}).
    \item \verb+model+: Definition of the \verb+AtomisticModel+ (see Section~\ref{sec:model}).
    \item \verb+task+: Configuration of the \verb+AtomisticTask+, including model outputs, losses and optimizers.
    \item \verb+trainer+: Arguments to be passed to the PyTorch Lightning \verb+Trainer+.
    \item \verb+callbacks+: A list of callbacks to be passed to the \verb+Trainer+
    \item \verb+logger+:  Configuration of training metric loggers, such as Tensorboard~\cite{tensorflow2015-whitepaper} or Aim~\cite{aimui}.
    \item \verb+seed+: Sets the random seed for PyTorch and Pytorch Lightning.
\end{itemize}

The first lines of the configuration in Fig.~\ref{fig:qm9config} show how the \verb+model+ and \verb+data+ are overridden with the default config templates (\verb+nnp+ and \verb+qm9+) for the neural network potential and the QM9 dataset, respectively.
The data and model templates are further modified in lines 14-36 by specifying pre- and postprocessing transforms, output modules as well as model outputs with losses and metrics to be tracked.
The special key \verb+_target_+ enables automatic instantiating of objects.
For example, in the list of model outputs in Fig.~\ref{fig:qm9config} (lines~26-30), the \verb+Atomwise+ module is initialized with the given parameters.
Here, the target object is not restricted to be a SchNetPack module, but can also be a class provided by a third-party Python package.
This makes it straightforward to extend SchNetPack with custom layers, losses and models.

\subsection{Modifying configurations from command line}
We have shown above how a training run specified by the experiment config can be started from the command line.
Beyond that, the configuration of a run can also be directly modified using the CLI.
For example, training neural network potentials with a different representation and using a larger learning rate than the default can be achieved as follows:
\begin{small}
\begin{minted}[xleftmargin=0pt, numbersep=0pt, numbers=none, mathescape]{bash}
spktrain experiment=qm9_atomwise \
  model/representation=schnet globals.lr=1e-3
\end{minted}
\end{small}
Note that when setting config groups to a preconfigured template, a slash '/' is used, while when setting a value in the YAML, the dot "." is used to navigate the hierarchy.
Finally, custom \verb+experiment+ configs can be added by setting a user config directory:
\begin{small}
\begin{minted}[xleftmargin=0pt, numbersep=0pt, numbers=none, mathescape]{bash}
spktrain -config-dir=/configdir \
  experiment=my_experiment
\end{minted}
\end{small}
The config directory needs to follow the structure of the config, e.g. the experiment should be located at \verb+/configdir/experiment/my_experiment.yaml+.

\subsection{Extending SchNetPack}
The modular design of SchNetPack configurations allows for seamless extension of the framework with additional models and learning tasks, including their integration into the CLI.
For example, the data pipeline and training flow of the framework could be kept while changing the representation block of a neural network potential.
This enables a quick and standardized comparison of different approaches.
To this end, one may implement a small python package containing the network building blocks as well as a custom configuration file for instantiating the new representation.
The training can be started with \verb+spktrain+ using the familiar CLI. 
By providing Hydra with the location of the additional configuration files in the extension package, the representation can simply be switched to the new network as if it was part of SchNetPack.
This means that developers are able to build atomistic ML models on top of SchNetPack by exchanging only selected building blocks and avoiding as much boilerplate code as possible.
The example in Section~\ref{sec:cgschnet} shows that this is not only possible with neural network potentials but also with more complex atomistic learning tasks such as a generative model for molecules.

\section{Molecular dynamics simulations}\label{sec:md}

In addition to the neural network library, SchNetPack 2.0 contains the \verb+schnetpack.md+ code for carrying out molecular dynamics simulations.
This MD environment is structured in a modular way in order to facilitate development and interfacing with different ML potentials.
Since all routines are implemented in PyTorch, models based on \verb+AtomisticModel+ can be used with minimal communication overhead and full GPU capabilities.

\begin{figure}
\includegraphics[width=\columnwidth]{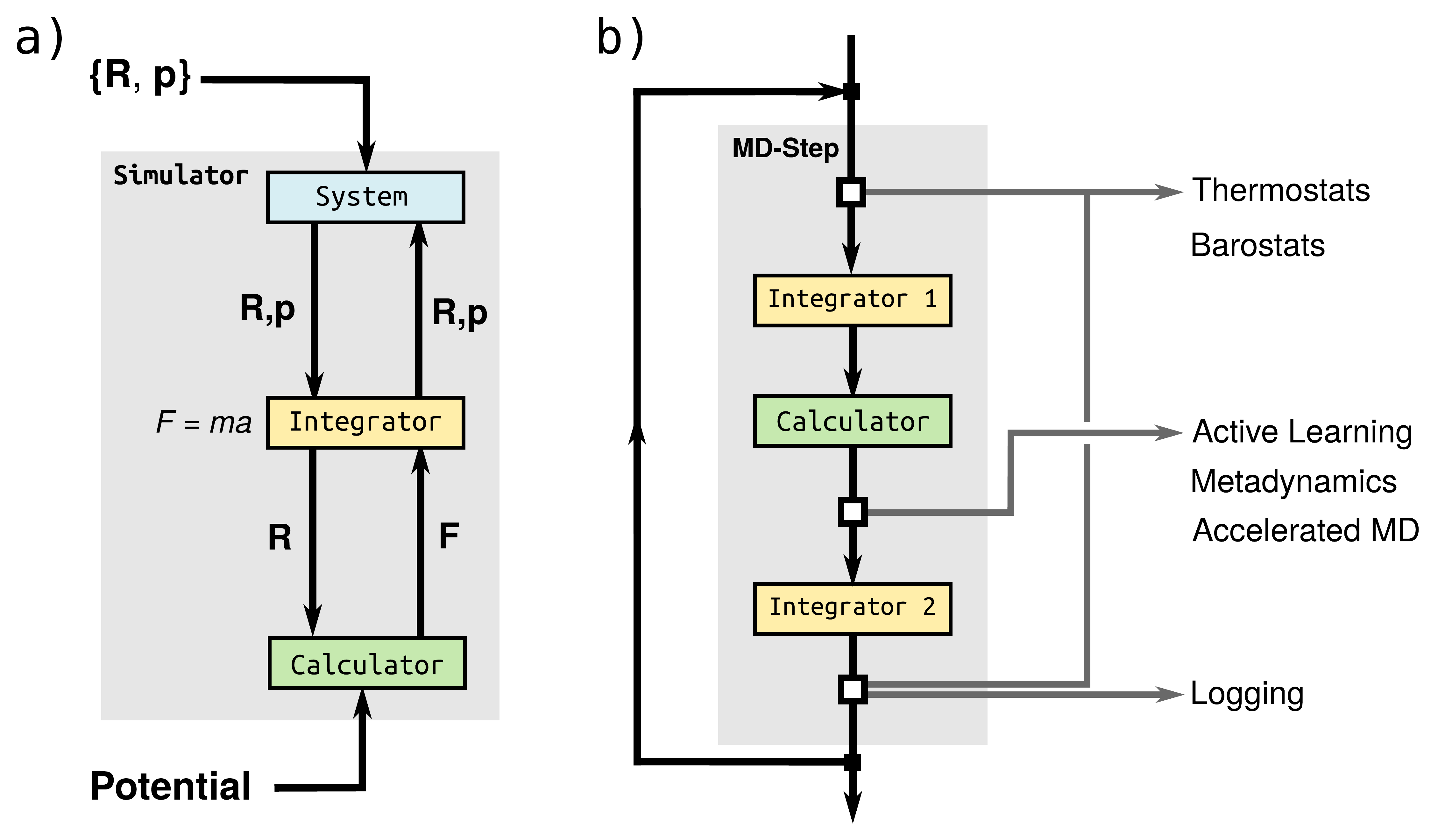}
\caption{\label{fig:md} a) Basic MD workflow. b) internal structure of a single simulation step, indicating points at which different hooks can be applied to modify a simulation.}
\end{figure}

The MD code performs several core tasks during each simulation step.
It keeps track of the positions $\mathbf{R}$ and momenta $\mathbf{p}$ of all nuclei, computes the forces $\mathbf{F}$ acting on them and uses the latter to integrate the equations of motion.
In SchNetPack, these tasks are distributed between different modules, which are sketched in Fig.~\ref{fig:md}a.
The \verb+md.System+ class contains all information on the current system state (e.g. nuclear positions and momenta).
The \verb+md.Integrator+ propagates the positions and momenta of the system.
In order to carry out this update, the nuclear forces are required.
These are computed by an \verb+md.Calculator+, which takes the positions of atoms and returns the forces due to the potential. 
Typically, the Calculator connects to a previously trained machine learning model.
All these modules are linked together in the \verb+md.Simulator+ class, which contains the main MD loop and calls the three previous modules.
The simulator can further be modified with so-called simulation hooks to control aspects like sampling (e.g. thermostats or barostats) or logging (Fig.~\ref{fig:md}b).
In the following, we will describe the MD modules in detail.

\subsection{System}\label{sec:system}

The \verb+md.System+ class keeps track of the state of the simulated system and collects all associated quantities (e.g. positions, momenta, atom types, simulation cells, \ldots).
This information is stored in multi-dimensional PyTorch tensors, which makes it possible to vectorize many operations and e.g. simulate several different molecules as well as different replicas of a molecule in a single step.
The shape convention of the system tensors is similar to the one used in the atomistic representations, with an added replica dimension $(n_\text{replicas}, n_\text{atoms}, \ldots)$.
This replica dimension collects different replicas of the same molecules and can be used, for example, in ring-polymer MD simulations.
In addition, the \verb+md.System+ class provides utilities for computing different quantities relevant for MD simulations, such as temperature and pressure.

Structural information can be added to a \verb+md.System+ instance via the \verb+load_molecules+ function using ASE \verb+Atoms+ objects.
The function operates on either a single object or a list of \verb+Atoms+ and automatically takes care of simulation cells and periodic boundary conditions.
If a list is provided, multiple molecules can be loaded at once.
During loading, it is necessary to specify the number of replicas and the units of length used for the input structures.
The latter is required, since the MD code uses its own internal unit system and automatically converts the inputs.
The creation of a \verb+md.System+ instance is demonstrated in the following example: 
\begin{small}
\begin{minted}[python3=true, xleftmargin=0pt, numbersep=0pt, numbers=none, mathescape]{python}
# Read structures with ASE
molecule = ase.io.read("molecule.xyz")

# Create system instance and load molecule
md_system = schnetpack.md.System()
md_system.load_molecules(
    molecule,
    n_replicas=1,
    position_unit_input="Angstrom"
)
\end{minted}
\end{small}

By default, momenta are set to zero upon creating a \verb+System+.
If desired, a \verb+md.Initializer+ can be used to draw the momenta from different distributions corresponding to a certain temperature. 
The following example uses the \verb+UniformInit+ initializer, which draws momenta from a random uniform distribution and rescales them to a certain temperature. 
Other routines, e.g. drawing from a Maxwell—Boltzmann distribution, are also available.

\begin{small}
\begin{minted}[python3=true, xleftmargin=0pt, numbersep=0pt, numbers=none, mathescape]{python}
md_initializer = md.UniformInit(
    300, # system temperature in K
    remove_center_of_mass=True,
    remove_translation=True,
    remove_rotation=True,
)

# Initialize the system momenta
md_initializer.initialize_system(md_system)
\end{minted}
\end{small}

A \verb+md.Initializer+ may also be used to center the molecular structure on the center of mass and remove all translational and rotational components of the momenta, as demonstrated in the code snippet above.

\subsection{Integrator}\label{sec:integrator}

As the name suggests, the \verb+md.Integrator+ is used to integrate the equations of motion based on the nuclear forces.
Two integration schemes are implemented in SchNetPack 2.0:
The first \verb+md.VelocityVerlet+ implements the Velocity Verlet algorithm which evolves the system in a purely classical manner.
The second integrator \verb+md.RingPolymer+ performs ring-polymer MD simulations, which recover a certain degree of nuclear quantum effects.
Both integrators use a three step scheme and come with special NPT variants for constant pressure simulations.

To initialize an \verb+md.Integrator+, one has to specify the integration time step in units of femtoseconds.
In the following example, we use the Velocity Verlet algorithm with a time step of $\Delta t=0.5$~fs: 
\begin{small}
\begin{minted}[python3=true, xleftmargin=0pt, numbersep=0pt, numbers=none, mathescape]{python}
time_step = 0.5 # fs

# Set up the integrator
md_integrator = md.VelocityVerlet(time_step)
\end{minted}
\end{small}

\subsection{Calculator}\label{sec:calculator}

The nuclear forces required for the integrator are provided by a \verb+md.Calculator+.
It serves as an interface between a computation method (e.g. the neural network potential) and the MD code.
The calculator takes the current positions of the \verb+md.System+ and other structural properties (simulation cells, etc) and converts them to a format suitable for the computation method.
Once all requested properties (e.g. forces) have been computed, the \verb+md.Calculator+ collects the results and reshapes the tensors back into the \verb+md.System+ format.
Different custom calculators can be derived from the \verb+MDCalculator+ base class.
SchNetPack comes with several predefined calculators, such as the \verb+SchNetPackCalculator+ which can be directly used with trained SchNetPack models, the \verb+OrcaCalculator+ for the ORCA quantum chemistry package\cite{neese2012WCMS}, as well as ensemble variants of these calculators for adaptive sampling.
In the following, we describe the basic usage of the \verb+SchNetPackCalculator+.

Since the core of the calculator is a SchNetPack \verb+AtomisticModel+, interatomic distances need to be computed during the MD.
This is done with the \verb+NeighborListMD+ wrapper, which can be used with any neighbor list transform (Tab.~\ref{tab:transforms}).
It takes the basic neighbor list module, a cutoff radius and a buffer region as input (both use the same length units as the model).
The introduction of a buffer region improves performance, since a MD neighbor list only needs to be recomputed once the structural changes exceeding the buffer region.
In the following, we use the \verb+MatScipyNeighborlist+, with a 5~{\AA} cutoff and 2~{\AA} buffer region:
\begin{small}
\begin{minted}[python3=true, xleftmargin=0pt, numbersep=0pt, numbers=none, mathescape]{python}
from schnetpack.md.neighborlist_md \ 
    import NeighborListMD
from schnetpack.transform \ 
    import MatScipyNeighborList

md_neighborlist = NeighborListMD(
    cutoff=5.0, # cutoff
    cutoff_shell=2.0, # buffer region
    MatScipyNeighborList,
)
\end{minted}
\end{small}

To initialize \verb+md.SchNetPackCalculator+, we need to provide it with the path to a trained model and the key of the forces in the output dictionary.
Moreover, the units that the calculator expects for positions and energy in order to properly convert between MD and calculator unit systems need to be specified. 
Force units are inferred automatically based on these two inputs.
Optionally, one may provide an \verb+energy_key+ to store the potential energies in the \verb+md.System+ class.
Additional properties (e.g. dipole moments for infrared spectra, etc.) can be requested with the \verb+required_properties+ argument.
A typical SchNetPack \verb+NeuralNetworkPotential+ trained on the MD17 data would be initialized as follows:
\begin{small}
\begin{minted}[python3=true, xleftmargin=0pt, numbersep=0pt, numbers=none, mathescape]{python}
from schnetpack.md.calculators \
    import SchNetPackCalculator

md_calculator = SchNetPackCalculator(
    "<PATH/TO/MODEL>",  # path to stored model
    "forces",  # force key
    "kcal/mol",  # energy units
    "Angstrom",  # length units
    md_neighborlist,  # neighbor list defined above
    energy_key="energy",  # potential energies
    required_properties=[],  # additional properties
)
\end{minted}
\end{small}

\subsection{Simulator}\label{sec:simulator}

The \verb+md.Simulator+ loops over a series of time steps, calls the individual MD modules in the appropriate order and updates the system state.
Its internal structure is depicted in Fig~\ref{fig:md}b.
Assuming a \verb+md.System+, \verb+md.Integrator+ and \verb+md.Calculator+ have been created as described above, the setup is straightforward:
\begin{small}
\begin{minted}[python3=true, xleftmargin=0pt, numbersep=0pt, numbers=none, mathescape]{python}
md_simulator = schnetpack.md.Simulator(
    md_system,
    md_integrator,
    md_calculator,
    simulation_hooks=[],
)
\end{minted}
\end{small}
Simulation hooks can be used to modify a simulation analogous to transforms in the neural network library and are described in Sec.~\ref{sec:mdhook}.
The required floating point precision and the computational device to be used can be set by the usual PyTorch directives:
\begin{small}
\begin{minted}[python3=true, xleftmargin=0pt, numbersep=0pt, numbers=none, mathescape]{python}
md_device = "cpu" 
md_precision = torch.float32

md_simulator = md_simulator.to(md_precision)
md_simulator = md_simulator.to(md_device)
\end{minted}
\end{small}

The \verb+simulate+ routine runs the simulation with the number of simulation steps as an argument:
\begin{small}
\begin{minted}[python3=true, xleftmargin=0pt, numbersep=0pt, numbers=none, mathescape]{python}
# simulate for 100 steps
md_simulator.simulate(100)
\end{minted}
\end{small}
Since the \verb+md.Simulator+ keeps track of the simulation and system state, repeated calls of \verb+calculate+ will resume the simulation.
 
\subsection{Simulation hooks}\label{sec:mdhook}

Simulation hooks can be used to tailor a simulation to specific needs, improving the customizability of the SchNetPack 2.0 MD code.
A \verb+SimulationHook+ can be thought of as a set of instructions for the \verb+Simulator+, which are performed at certain points of a MD step.
Depending on which time they are called and which actions they encode, simulation hooks can achieve a wide range of tasks.

Fig~\ref{fig:md}b shows an overview at which points simulation hooks can be applied during a \verb+Simulator+ step. 
If a \verb+SimulationHook+ is applied before and after the integration half-step updating the momenta, it could control temperature and pressure of the system in the form of thermostats and barostats.
Acting directly after the \verb+md.Calculator+, a hook may implement custom sampling schemes such as metadynamics or adaptive sampling.\cite{laio2002escaping,gastegger2017machine}
When called after the second integration step, simulation hooks can be used to collect and store information on the system for analysis.
A list of multiple hooks can be passed to a simulator, which enables control of a simulation in various manners. 
An overview of all simulation hooks currently implemented in SchNetPack 2.0 is shown in Tab.~\ref{tab:hooks}.
In the following, we will give a brief overview on how to apply thermostat hooks and collect simulation data:

\begin{table*}
\renewcommand\arraystretch{1.2}
\caption{\label{tab:hooks} Overview of simulation hooks (and data streams).}
\begin{ruledtabular}
\begin{tabular}{llp{0.55\linewidth}}
\textbf{\textit{Category}} & \textbf{\textit{Hook}} & \textbf{\textit{Description}}\\
\hline
\textbf{Basic}       & \verb+SimulationHook+           & Abstract base class for deriving hooks. \\
                     & \verb+RemoveCOMMotion+          & Periodically remove center of mass translation and rotation. \\
\hline
\textbf{Thermostats} & \verb+BerendsenThermostat+      & Simple velocity rescaling thermostat.\cite{berendsen1984molecular} \\
                     & \verb+LangevinThermostat+       & Basic stochastic Langevin thermostat.\cite{bussi2007accurate} \\
                     & \verb+NHCThermostat+            & Nos{\'e}--Hoover chain thermostat.\cite{martyna1992nose} \\
                     & \verb+GLEThermostat+            & Stochastic generalized Langevin equantion (GLE) colored nose thermostat.\cite{ceriotti2010colored} \\
                     & \verb+PILELocalThermostat+      & Path integral Langevin equation (PILE) thermostat for ring-polymer MD.\cite{ceriotti2010efficient} \\
                     & \verb+PILEGlobalThermostat+     & Global variant of PILE, applies stochastic velocity rescaling to the ring-polymer centroid.\cite{ceriotti2010efficient} \\
                     & \verb+TRPMDThermostat+          & Thermostated ring-polymer variant of the PILE thermostat.\cite{rossi2014remove} \\
                     & \verb+RPMDGLEThermostat+        & GLE colored noise thermostat for ring-polymer MD.\cite{ceriotti2010colored} \\
                     & \verb+PIGLETThermostat+         & Version of GLE where every normal mode is thermostated seperately.\cite{uhl2016accelerated} \\
                     & \verb+NHCRingPolymerThermostat+ & Nos{\'e}--Hoover chain thermostat for ring-polymer MD.\cite{ceriotti2010efficient} \\
\hline
\textbf{Barostats}   & \verb+NHCBarostatIsotropic+     & Combined Nos{\'e}--Hoover chain thermostat and barostat for isotropic cell fluctuations.\cite{martyna1996explicit} \\
                     & \verb+NHCBarostatAnisotropic+   & Combined Nos{\'e}--Hoover chain thermostat and barostat for anisotropic cell fluctuations.\cite{martyna1996explicit} \\
                     & \verb+PILEbarostat+             & Stochastic PILE barostat for ring-polymer MD.\cite{kapil2019pi} \\
\hline
\textbf{Logging}     & \verb+CheckPoint+               & Periodically store the system state for restarting. \\
                     & \verb+TensorBoardLogger+         & Log system information (e.g. temperature, energy) in TensorBoard format. \\
                     & \verb+FileLogger+               & Log system information to a custom HDF5 dataset. Data streams are used to store different data groups. \\
                     & \verb+MoleculeStream+           & Data stream for storing structural information with the \verb+FileLogger+. \\ 
                     & \verb+PropertyStream+           & Data stream for storing system properties with the \verb+FileLogger+. \\ 
\end{tabular}
\end{ruledtabular}
\end{table*}

\paragraph{Temperature and pressure control}
Constant temperature (NVT) and constant pressure (NPT) simulations can be performed using thermostat and barostat hooks.
A simple example is the stochastic Langevin thermostat implemented in \verb+LangevinThermostat+.
It requires a time constant (in fs) and bath temperature (in K) to initialize:
\begin{small}
\begin{minted}[python3=true, xleftmargin=0pt, numbersep=0pt, numbers=none, mathescape]{python}
from schnetpack.md.simulation_hooks \
    import LangevinThermostat
    
thermostat_hook = LangevinThermostat(
    300.0, # bath temperature in K
    100.0, # time constant in fs
)
\end{minted}
\end{small}

\paragraph{Callbacks and Logging} 
The \verb+FileLogger+ hook is the primary way to collect and store MD simulation data.
The kind of information stored is controlled via two types of data streams:
The \verb+MoleculeStream+ stores structural information and velocities during the simulation, while \verb+PropertyStream+ collects properties computed by the \verb+md.Calculator+.
To reduce file I/O overhead, the \verb+FileLogger+ collects a certain number of steps into a buffer before it is written to an HDF5 file at once.

The \verb+FileLogger+ requires the destination path, the buffer size and the data streams. In addition, the logging frequency can be specified:
\begin{small}
\begin{minted}[python3=true, xleftmargin=0pt, numbersep=0pt, numbers=none, mathescape]{python}
from schnetpack.md.simulation_hooks \
    import callback_hooks

# Set up data streams
data_streams = [
    # store positions and velocities
    callback_hooks.MoleculeStream(
        store_velocities=True
    ),
    # store energies
    callback_hooks.PropertyStream(
        target_properties=["energy"]
    ),
]

# Create the file logger
file_logger_hook = callback_hooks.FileLogger(
    "simulation.hdf5", # path to the log file
    100,               # size of the buffer
    data_streams=data_streams,
    every_n_steps=1,  # logging frequency
)
\end{minted}
\end{small}

\subsection{Using the HDF5 dataset}\label{sec:mdhdf5}

SchNetPack 2.0 simulation data, stored in a HDF5 dataset as described previously, can be accessed with the \verb+HDF5Loader+ class to retrieve structures or perform analysis.
Properties collected during simulation can be extracted with the \verb+get_property+ function.
It takes the name of a property and an indicator, whether the property relates to the whole system or particular atoms:
\begin{small}
\begin{minted}[python3=true, xleftmargin=0pt, numbersep=0pt, numbers=none, mathescape]{python}
from schnetpack.md.data \
    import HDF5Loader

md_data = HDF5Loader("simulation.hdf5")

energy = md_data.get_property(
    "energy", atomistic=False
)
\end{minted}
\end{small}
The \verb+HDF5Loader+ comes with multiple predefined functions to extract specific properties, such as temperature or pressure.
All of those receive a molecule index \verb+mol_idx+ and replica index \verb+repica_idx+ as additional inputs.
The former is used to extract information of a particular system when multiple are simulated at the same time, while the latter is used to target specific replicas, e.g. in ring-polymer MD.
The default behavior is to target the first system and compute the average (centroid) over all replicas.
Finally, \verb+convert_to_atoms+ allows structures to be extracted as a list of ASE \verb+Atoms+ objects.

SchNetPack 2.0 provides the module \verb+md.data.spectra+ to compute different vibrational spectra directly from an \verb+HDF5Loader+ instance.
Routines for power spectra (\verb+PowerSpectrum+, requires momenta), infrared (IR) spectra (\verb+IRSpectrum+, requires dipole moments) and Raman spectra (\verb+RamanSpectrum+, requires polarizabilities) are available.
For example, the power spectrum can be computed as follows:
\begin{small}
\begin{minted}[python3=true, xleftmargin=0pt, numbersep=0pt, numbers=none, mathescape]{python}
from schnetpack.md.data \ 
    import PowerSpectrum

# Initialize the spectrum
spectrum = PowerSpectrum(md_data, resolution=2048)

# Compute the spectrum
spectrum.compute_spectrum()

# Get frequencies and intensities
frequencies, intensities = spectrum.get_spectrum()
\end{minted}
\end{small}

\subsection{Molecular Dynamics Configuration and Command Line Interface}

Similar to the neural network package, MD simulations can be configured using the Hydra framework.
The central command of the SchNetPack MD CLI is \verb+spkmd+, which sets up everything for a basic MD simulation and creates a simulation directory.
Runs can be configured via predefined config groups and command-line overrides.
Moreover, since the Hydra CLI is able to instantiate classes from YAML configs, it is straightforward to integrate external modules, such as custom calculators for simulations.

A standard \verb+spkmd+ run requires the following inputs:
\begin{small}
\begin{minted}[xleftmargin=0pt, numbersep=0pt, numbers=none, mathescape,breaklines]{bash}
spkmd simulation_dir=<DIR> system.molecule_file=<XYZ> calculator.model_file=<MODEL> calculator.neighbor_list.cutoff=<CUTOFF>
\end{minted}
\end{small}
where \verb+simulation_dir+ indicates the simulation directory, \verb+system.molecule_file+ the file containing the structures to be simulated and \verb+calculator.model_file+ and \verb+calculator.neighbor_list.cutoff+ specify the path to the previously trained neural network potential and the cutoff radius used in the model.
This starts a MD run with a predefined default configuration, which corresponds to a NVE simulation where features such as logging and checkpointing have already been set up.



Like in the neural network package, an MD config is structured into different config groups:
\begin{itemize}
    \item uncategorized, general settings, such as device, precision, random seed and simulation directory.
    \item \verb+calculator+: Definition of the MD calculator (see Sec.~\ref{sec:calculator}).
    \item \verb+system+: Definition of the MD system, including loading of structures and initial conditions (see Sec.~\ref{sec:system}).
    \item \verb+dynamics+: Definition of the overall MD loop (Sec.~\ref{sec:simulator}). Contains the subgroups:
    \begin{itemize}
        \item \verb+integrator+: Integrator settings (Sec.~\ref{sec:integrator})
        \item \verb+thermostat+: Temperature control.
        \item \verb+barostat+: Pressure control.
        \item \verb+simulation_hook+: General hooks for sampling (Sec.~\ref{sec:mdhook}).
    \end{itemize}
    \item \verb+callbacks+: Definition of hooks for callback and logging.
\end{itemize}

These groups can be used to further configure the simulation, e.g. by adding a thermostat or changing integrator settings (see e.g. Sec~\ref{sec:fieldschnet}). 

MD simulations can be started using full or partial config files in YAML format as input.
For example, it is possible to create a basic config file by calling \verb+spkmd+ with the \verb+--cfg job+ flag and edit it for a particular application.
An existing config file can then be loaded with the \verb+load_config+ option:
\begin{small}
\begin{minted}[xleftmargin=0pt, numbersep=0pt, numbers=none, mathescape,breaklines]{bash}
spkmd simulation_dir=<DIR> load_config=<CONFIG-FILE>
\end{minted}
\end{small}
It is still possible to further modify the simulation via other command line overwrites.

\section{Example applications}\label{sec:examples}

This section features some basic and advanced applications of SchNetPack.
First, we demonstrate how to train atomistic neural networks on the supported benchmark datasets.
In a second example, we showcase the prediction of response properties at the example of a custom dataset.
Finally, we demonstrate how SchNetPack can be extended with custom code to train a generative neural network for the inverse design of 3d structures.

\subsection{Potential energy surfaces and property prediction}\label{sec:qm9md17}
\begin{table*}
\renewcommand\arraystretch{1.2}
\caption{\label{tab:qm9md17} Mean absolute and mean squared error for various neural networks trained with SchNetPack on prediction task on the QM9 and rMD17 benchmark datasets. The errors are averaged over three repetitions.}
\begin{ruledtabular}
\begin{tabular}{llllrrrr}
\textbf{Dataset} & \textbf{Property} & \textbf{Unit} & \textbf{Metric} &
\textbf{SchNet} & \textbf{PaiNN} & \textbf{SO3net} (l$_\text{max}$=1) & \textbf{SO3net} (l$_\text{max}$=2) \\
\hline
\multirow{2}{*}{QM9} & \multirow{2}{*}{U$_0$} & \multirow{2}{*}{meV} & MAE & 9.6 & 5.7 & 6.8 & 6.4 \\
&  &  & RMSE & 21.9 & 15.3 & 16.2 & 17.1 \\ \hline
\multirow{2}{*}{QM9}  & \multirow{2}{*}{$\mu$} & \multirow{2}{*}{Debye} & MAE & 0.022 & 0.011 & 0.018 & 0.014 \\
&  &  & RMSE & 0.044 & 0.026 & 0.038 & 0.033 \\ \hline
\multirow{4}{*}{rMD17 / Aspirin}  & \multirow{2}{*}{$E$} & \multirow{2}{*}{meV} & MAE & 13.5 & 3.8 & 3.8 & 2.6 \\
&  &  & RMSE & 18.3 & 5.9 & 5.7 & 3.8 \\ 
& \multirow{2}{*}{$F$} & \multirow{2}{*}{meV/{\AA}} & MAE & 33.2 & 12.8 & 12.7 & 9.0 \\
&  &  & RMSE & 49.5 & 21.7 & 19.6 & 14.5 \\ \hline
\multirow{4}{*}{rMD17 / Paracetamol}  & \multirow{2}{*}{$E$} & \multirow{2}{*}{meV} & MAE & 8.4 & 2.1 & 2.2 & 1.4 \\
&  &  & RMSE  & 11.2 & 2.9 & 3.0 & 1.9 \\ 
& \multirow{2}{*}{$F$} & \multirow{2}{*}{meV/{\AA}} & MAE & 26.1 & 9.0 & 8.9 & 6.0 \\
&  &  & RMSE & 40.0 & 14.7 & 13.8 & 10.0
\end{tabular}
\end{ruledtabular}
\end{table*}

\begin{table}
\renewcommand\arraystretch{1.2}
\caption{\label{tab:timing} Training and validation time per epoch for models trained on QM9 and rMD17 tasks using an Nvidia A100.}
\begin{ruledtabular}
\begin{tabular}{llrr}
\textbf{Task} & \textbf{Model}  & \textbf{\# params} &  \textbf{time / epoch} \\ \hline
\multirow{4}{*}{\parbox{1.7cm}{QM9, U$_0$ \\ 110k / 10k \\ bs 100}} & SchNet & 432k & 1 min 14 sec \\
& PaiNN & 589k & 1 min 13 sec \\
& SO3net (l$_\text{max}$=1) & 283k & 1 min 16 sec \\
& SO3net (l$_\text{max}$=2) & 341k & 2 min 29 sec \\ \hline
\multirow{4}{*}{\parbox{1.7cm}{rMD17, E+F, Aspirin \\ 950 / 50 \\ bs 10}} & SchNet & 432k & 7 sec \\
& PaiNN & 589k & 6 sec \\
& SO3net (l$_\text{max}$=1) & 283k & 7 sec \\
& SO3net (l$_\text{max}$=2) & 341k & 10 sec
\end{tabular}
\end{ruledtabular}
\end{table}

The datasets QM9~\cite{qm9two} and MD17~\cite{chmiela2017machine} have become established benchmarks in the development of atomistic representations.
Here, we use the revised MD17 (rMD17) dataset for which the energies and forces have been recomputed with higher accuracy~\cite{christensen2020role}.
We have trained models using three representations SchNet, PaiNN and SO3net.
For the latter, we have explored setting the maximum angular moment to $l_\text{max} \in \{ 1, 2\}$.

We have predicted the inner energy $U_0$ and the total dipole moment $\mu$ from the properties in QM9 with two separate models.
The configuration \verb+qm9_atomwise+ for $U_0$ is given in Fig.~\ref{fig:qm9config}.
The only difference of the \verb+qm9_dipole+ configuration is the use of the \verb+DipoleMoment+ output module:
\begin{scriptsize}
\begin{minted}[xleftmargin=0pt, numbersep=0pt, numbers=none, mathescape]{yaml}
model:
  output_modules:
    - _target_: schnetpack.atomistic.DipoleMoment
      dipole_key: ${globals.property}
      n_in: ${model.representation.n_atom_basis}
      predict_magnitude: True
      use_vector_representation: False
  postprocessors:
    - _target_: schnetpack.transform.CastTo64
\end{minted}
\end{scriptsize}
If the representation supports equivariant vector features, which is the case for PaiNN and SO3net, we use atomic dipoles in the output layer~\cite{schutt2021equivariant} by setting \verb+use_vector_representation=True+ (see Table~\ref{tab:atomistic}).

The training runs have been started using the SchNetPack CLI.
The initial learning rate was set such that training was still stable, which we found to be $5 \times 10^{-4}$ for SchNet and PaiNN and $1 \times 10^{-3}$ for SO3net.
We use a learning rate scheduler that decays when the validation error does not decrease within the given number of patience epochs.
The training of dipole moments can be reproduced by calling:
\begin{small}
\begin{minted}[xleftmargin=0pt, numbersep=0pt, numbers=none, mathescape,breaklines]{bash}
spktrain experiment=qm9_dipole task.scheduler_args.patience=25 trainer.max_epochs=5000 model/representation/radial_basis=bessel task.optimizer_args.weight_decay=0.01 globals.lr={1e-3,5e-4} model/representation={schnet,painn,so3net} [model.representation.lmax={1,2}] model.output_modules.0. use_vector_representation={True,False}
\end{minted}
\end{small}
Different hyperparameter selections are indicated by curly brackets and \verb+lmax+ can only be set for SO3net ( indicated by \verb+[]+).


We have selected the aspirin and paracetamol molecules for our rMD17 experiments and predicted energies and forces.
We call the \verb+md17+ experiment configuration of SchNetPack with the following settings:
\begin{small}
\begin{minted}[xleftmargin=0pt, numbersep=0pt, numbers=none, mathescape,breaklines]{bash}
spktrain experiment=md17 data=rmd17 data.molecule={aspirin,paracetamol} globals.lr={1e-3,5e-4}  task.optimizer_args.weight_decay=0.01 model/representation={schnet,painn,so3net} [model.representation.lmax={1,2}]
\end{minted}
\end{small}

Table~\ref{tab:qm9md17} shows the the errors for the trained SchNetPack models. 
The results for SchNet and PaiNN are similar to what has been observed in earlier work~\cite{schutt2018schnet,schutt2021equivariant}, although hyperparameters such as the learning rate schedule, the radial basis and the weight decay may differ.
Note, that these experiment are only meant to demonstrate the capabilities of SchNetPack.
For an accurate comparison between atomistic ML models, an extensive hyperparameter search should be performed.
Table~\ref{tab:timing} shows the average time per epoch of the performed experiments. 
Even though SchNet and PaiNN have more parameters than SO3net, this does not correspond directly to the number of floating point operations.
In particular, when setting the maximum angular momentum $l_\text{max}>1$, the required compute rises faster than the model size due to increased parameter sharing in our tensor product layer implementation.

\subsection{Modeling response properties \label{sec:fieldschnet}}

The FieldSchNet representation\cite{gastegger2021machine}, included in SchNetPack 2.0, is able to model response properties and solvent effects.
Here, we demonstrate how to train a FieldSchNet model for potential energies $E$ and corresponding response properties, namely atomic forces $\mathbf{F}$, dipole moments $\boldsymbol{\mu}$, polarizabilities $\boldsymbol{\alpha}$ and nuclear shielding tensors $\boldsymbol{\sigma}$.
We use the reference data of ethanol in vacuum published in Ref.~\citenum{gastegger2017machine}.

Two modifications to the standard neural network potential configuration are required:
First, a \verb+StaticExternalFields+ module needs to be added to the input modules to set up the required auxiliary fields (electric and magnetic).
Second, a \verb+Response+ output module is appended after the \verb+Atomwise+ layer in order to compute the different derivatives corresponding to the response properties.
Both modules automatically determine the required fields and derivatives based on the requested response properties (in this case specified in \verb+globals.response_properties+):
\begin{scriptsize}
\begin{minted}[xleftmargin=0pt, numbersep=0pt, numbers=none, mathescape]{yaml}
globals:
  energy_key: energy
  response_properties:
    - forces
    - dipole_moment
    - polarizability
    - shielding

model:
  input_modules:
    - _target_: schnetpack.atomistic.PairwiseDistances
    - _target_: schnetpack.atomistic.StaticExternalFields
      response_properties: ${globals.response_properties}
  output_modules:
    - _target_: schnetpack.atomistic.Atomwise
      output_key: ${globals.energy_key}
      n_in: ${model.representation.n_atom_basis}
      aggregation_mode: sum
    - _target_: schnetpack.transform.ScaleProperty
      input_key: ${globals.energy_key}
      output_key: ${globals.energy_key}
    - _target_: schnetpack.atomistic.Response
      energy_key: ${globals.energy_key}
      response_properties: ${globals.response_properties}
\end{minted}
\end{scriptsize}

These changes and settings such as loss, metrics and tradeoffs are predefined in the experiment config \verb+response+, which only requires the dataset, batch size and splits to be specified:
\begin{small}
\begin{minted}[xleftmargin=0pt, numbersep=0pt, numbers=none, mathescape,breaklines]{bash}
spktrain experiment=response data.datapath=<PATH/TO/DB> data.num_train=8000 data.num_val=1000 data.batch_size=20
\end{minted}
\end{small}
This command trains a FieldSchNet model with a cutoff of 9.449~Bohr (5~{\AA}), 128 features and five interactions, using 9\,000 points of the ethanol dataset for training and validation and the remaining 1\,000 points for testing.
The average test errors over three runs can be found in Tab.~\ref{tab:ethanolfield}.
\begin{table}
	\caption{Prediction error of the FieldSchNet model for the ethanol molecule.}
	\label{tab:ethanolfield}
    \begin{ruledtabular}
	\begin{tabular}{llr}
		\textbf{Property}     & \textbf{Unit}                 & MAE   \\
		\hline
		$\mathbf{E}$          & kcal\,mol$^{-1}$              &   0.023 \\ 
		$\mathbf{F}$          & kcal\,mol$^{-1}$\,{\AA}$^{-1}$&   0.158 \\ 
		$\boldsymbol{\mu}$    & D                             &   0.004 \\ 
		$\boldsymbol{\alpha}$ & Bohr$^3$                      &   0.009 \\ 
		$\sigma_{H}$          & ppm                           &   0.045 \\ 
		$\sigma_{C}$          & ppm                           &   0.215 \\ 
		$\sigma_{O}$          & ppm                           &   0.469 \\ 
	\end{tabular}
    \end{ruledtabular}
\end{table}

Once the model is trained, we can use the following \verb+spkmd+ command to perform a 25~ps ring-polymer simulation with 16 beads:
\begin{small}
\begin{minted}[xleftmargin=0pt, numbersep=0pt, numbers=none, mathescape,breaklines]{bash}
spkmd simulation_dir=<DIR> system.molecule_file=<ethanol.xyz> calculator.model_file=<MODEL> calculator.neighbor_list.cutoff=9.449 calculator.energy_unit=Hartree calculator.position_unit=Bohr system.n_replicas=16 dynamics/integrator=rpmd +dynamics/thermostat=pi_nhc_global dynamics.n_steps=125000  calculator.required_properties=[ energy, dipole_moment, polarizability ] callbacks.hdf5.data_streams.1.target_properties=[ energy, dipole_moment, polarizability ]
\end{minted}
\end{small}
Since the reference data uses atomic units, we set the energy and position units in the calculator to Hartree and Bohr.
The simulation temperature is kept at 300~K with a ring-polymer Nos{\'e}--Hoover chain thermostat (see~\ref{tab:hooks}).
The overrides \verb+calculator.required_properties+ and \verb+callbacks.hdf5.data_streams.1.target_properties+ provide instructions which properties the \verb+md.Calculator+ needs to compute and which ones should be logged by the \verb+FileLogger+.
In this case, we select the potential energy, dipole moment and polarizability in order to compute vibrational infrared and Raman spectra.

\begin{figure}
\includegraphics[width=\columnwidth]{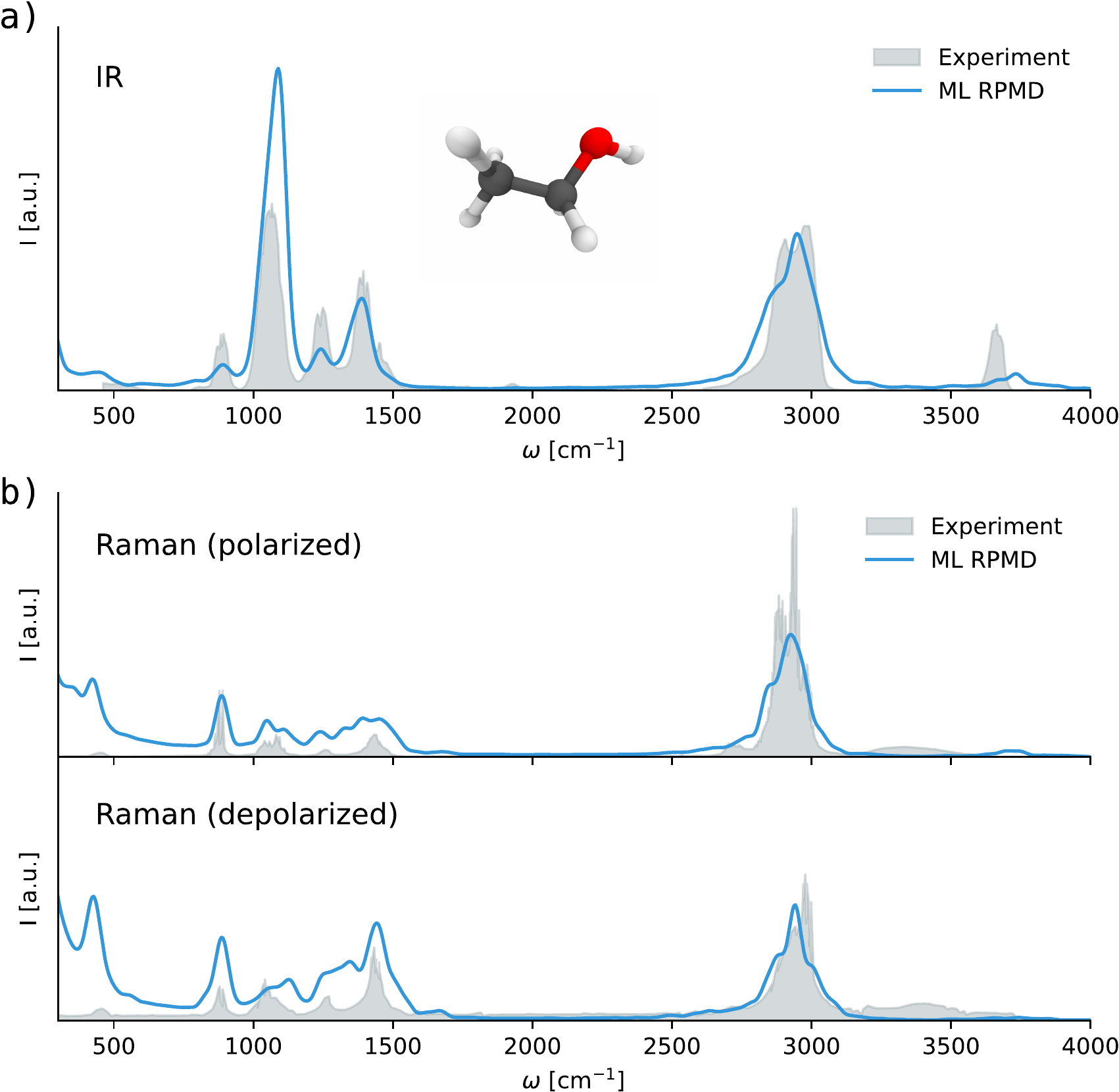}
\caption{\label{fig:mdspectra} Vibrational spectra of ethanol obtained by a ring-polymer MD using a FieldSchNet model. Shown are (a) the infrared and (b) polarized and depolarized Raman spectra along with respective experimental references (gray).}
\end{figure}

The \verb+HDF5Loader+ is used to load the \verb+simulation.hdf5+ file generated by the MD, where we skip an initial equilibration period of 5~ps (25\,000 steps).
Finally, we compute the infrared and Raman spectra with the \verb+IRSpectrum+ and \verb+RamanSpectrum+ routines, using a temperature of 300~K and laser frequency of 514~nm for the latter.
The predicted and experimental spectra can be found in Fig.~\ref{fig:mdspectra}.

\subsection{Building generative models}\label{sec:cgschnet}

\begin{figure*}
\raggedright
\begin{minipage}[t]{0.24\textwidth}
\raggedright
\textbf{a}
\begin{tiny}
\DTsetlength{0.2em}{1.0em}{0.2em}{0.1pt}{1.4pt}
\setlength{\DTbaselineskip}{10pt}
\dirtree{%
.1 \texttt{configs/}.
.2 \texttt{data/}.
.3 \texttt{gschnet\_qm9.yaml}.
.2 \texttt{experiment/}.
.3 \texttt{gschnet\_qm9\_gap\_relenergy.yaml}.
.2 \texttt{model/}.
.3 \texttt{conditioning/}.
.4 \texttt{gap\_relenergy.yaml}.
.3 \texttt{gschnet.yaml}.
.2 \texttt{task/}.
.3 \texttt{gschnet\_task.yaml}.
.2 \texttt{\_\_init\_\_.py}.
}
\end{tiny}
\end{minipage}
\begin{minipage}[t]{0.4\textwidth}
\raggedright
\textbf{b}
\begin{tiny}
\renewcommand\theFancyVerbLine{%
\ifnum\value{FancyVerbLine}=6
  \setcounter{FancyVerbLine}{44}\ldots
\else\ifnum\value{FancyVerbLine}=49
  \setcounter{FancyVerbLine}{64}\ldots
\else\ifnum\value{FancyVerbLine}=73
  \setcounter{FancyVerbLine}{100}\ldots
\else
\arabic{FancyVerbLine}%
\fi
\fi
\fi
}
\begin{minted}[xleftmargin=10pt, numbersep=5pt, mathescape, linenos, escapeinside=||]{yaml}
defaults:
 - override /data: gschnet_qm9
 - override /task: gschnet_task
 - override /model: gschnet
 - override /model/conditioning: gap_relenergy

data:
 transforms:
  - _target_: schnetpack.transform.SubtractCenterOfMass
  - _target_: schnetpack_gschnet.transform.OrderByDistanceToOrigin
  
  - _target_: schnetpack_gschnet.transform.BuildAtomsTrajectory
   centered: True
   origin_type: ${globals.origin_type}
   focus_type: ${globals.focus_type}
   stop_type: ${globals.stop_type}
   draw_random_samples: ${globals.draw_random_samples}
   sort_idx_i: False
  - _target_: schnetpack.transform.CastTo32
|\phantom|
\end{minted}
\end{tiny}
\end{minipage}
\begin{minipage}[t]{0.32\textwidth}
\raggedright
\textbf{c}
\includegraphics[width=1.0\textwidth]{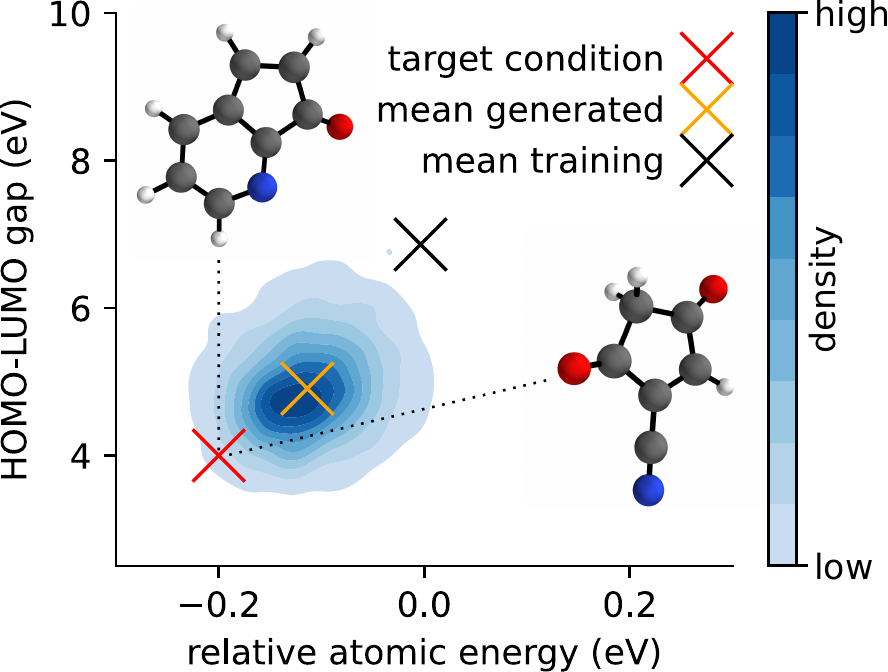}
\end{minipage}
\caption{\textbf{a} Additional config files from \texttt{schnetpack-gschnet}. The directory tree follows the structure induced by SchNetPack 2.0 in order to enable the composition of \texttt{schnetpack} and \texttt{schnetpack-gschnet} config files. \textbf{b} Excerpt from the experiment config \texttt{gschnet\_qm9\_gap\_relenergy.yaml} where we override defaults from \texttt{schnetpack} with the new config files and use transforms from both packages in order to train a cG-SchNet model conditioned on HOMO-LUMO gap and energy of molecules on the QM9 data set. \textbf{c} Density plot showing the HOMO-LUMO gap and energy of 20k molecules generated after training with \texttt{schnetpack-gschnet}. We use particularly low values of gap and energy as target (red cross) and show the two generated molecules closest to it. The mean gap and energy of training structures (black cross) and generated structures (orange cross) are also provided. Energy and gap of generated structures are predicted with PaiNN models trained with standard settings from SchNetPack 2.0.\label{fig:cg-schnet}}
\end{figure*}

Generative SchNet (G-SchNet)~\cite{gebauer2019gschnet} is a deep autoregressive neural network model for the inverse design of 3d molecular structures.
Recently, the model has been extended to learn conditional distributions by taking target property values as additional inputs~\cite{gebauer2022inverse}.
We have implemented an updated version of this conditional G-SchNet (cG-SchNet) as an extension of SchNetPack 2.0.
The package is called \verb+schnetpack-gschnet+ and available on Github\footnote{{https://github.com/atomistic-machine-learning/schnetpack-gschnet}}.
Compared to previous implementations, it aims at simple integration of custom data sets and improves the scalability of cG-SchNet both in terms of memory and computational complexity in order to make it applicable to larger molecules.

Since cG-SchNet is not a neural network potential but a generative model, several new modules are required for the architecture, the atomistic task, and the data processing.
The core network implementation consists of the classes \verb+ConditionalGenerativeSchNet+, a subclass of \verb+AtomisticModel+, and \verb+ConditioningModule+, which takes any amount of \verb+ConditionEmbedding+ networks to extract a combined feature vector representing all conditioning targets.
Three subclasses of \verb+ConditionEmbedding+ are provided for the embedding of scalar properties, vectorial properties, and the atomic composition of molecules.
Furthermore, \verb+ConditionalGenerativeSchNetTask+, a subclass of \verb+AtomisticTask+, customizes the learning task including the loss functions applied to predicted distributions and some task-specific setup, e.g. making sure that the molecular properties required as target conditions are loaded by the data module.
To learn the sequential placements of atoms with cG-SchNet, training molecules need to be sliced into trajectories where the structure grows atom by atom.
This is implemented in a preprocessing \verb+Transform+ called \verb+BuildAtomsTrajectory+, which allows to sample a random trajectory for each data point in each epoch.
The process depends on a few prerequisites, e.g. a certain ordering of atoms and different neighbor lists, which are also computed in custom transforms.
Additionally, we require a filter to exclude disconnected structures which is evaluated once prior to determining the training, validation, and test splits.
Thus, it is not implemented as a transform but in the setup stage of \verb+GenerativeAtomsDataModule+, which is a subclass of \verb+AtomsDataModule+ and serves as the base class for data sets used with cG-SchNet.
The package contains \verb+QM9Gen+, an example data set class for the QM9 benchmark data set.

The hierarchical configuration framework Hydra allows to easily integrate the new modules with SchNetPack.
It requires corresponding YAML files for the model, task, data, and experiment, where the directory tree should follow the config groups of SchNetPack as described in Section~\ref{sec:cli} (see Fig.~\ref{fig:cg-schnet}a).
We start the training of cG-SchNet just like for other models via the SchNetPack CLI by supplying the new configs as additional sources:
\begin{small}
\begin{minted}[xleftmargin=0pt, numbersep=0pt, numbers=none, mathescape, breaklines]{bash}
spktrain --config-dir=<PATH/TO/CONFIGS> experiment=gschnet_qm9_gap_relenergy
\end{minted}
\end{small}
Here, we use a custom experiment that overrides the model, task, and data configs to train a cG-SchNet model on QM9 that is conditioned on the energy and the HOMO-LUMO gap.
All remaining configs, e.g. for the trainer and the run, are loaded from SchNetPack.
In Fig.~\ref{fig:cg-schnet}b, we show the integration of new configs and how \verb+Transform+ modules from both \verb+schnetpack+ and \verb+schnetpack-gschnet+ can be accessed in the experiment configuration.

While the training reuses code and configs from the SchNetPack framework, the inference with cG-SchNet consists of sampling molecules from scratch, which is quite different from predicting properties of given molecules.
Therefore, the generation of molecules is implemented in the package as a separate CLI with its own, hierarchical Hydra configuration.
Exemplary results of generated molecules with cG-SchNet after training with our SchNetPack extension are shown in Fig.~\ref{fig:cg-schnet}c.

\section{Conclusions}

We have presented SchNetPack 2.0 which constitutes a major upgrade in 
functionality over the first version.
The new data pipeline comes with preprocessing transforms and a sparse data format.
Due to precalculated indices, sparse operations within the model, such as aggregation of neighbors in message passing or Clebsch-Gordan tensor products can be written in a couple of lines.
The switch to versatile training and configuration frameworks makes it easy for developers to extend SchNetPack with custom modules, datasets and configs.
SchNetPack 2.0 moves beyond neural network potentials by enabling a flexible definition of complex training tasks, as we have shown at the example of a generative neural network for 3d molecules.
Finally, SchNetPack comes with its own molecular dynamics simulation code so that trained models can directly be applied.
We are confident that these changes and additions in SchNetPack 2.0 will prove useful for both users and developers of atomistic neural networks.

\begin{acknowledgments}
We thank all contributors to SchNetPack on Github.
KTS, SSPH, NWAG and JL acknowledge support by the Federal Ministry of Education and Research (BMBF) for the Berlin Institute for the Foundations of Learning and Data (BIFOLD) (01IS18037A).
MG and NWAG work at the BASLEARN – TU Berlin/BASF Joint Lab for Machine Learning, co-financed by TU Berlin and BASF SE.
\end{acknowledgments}

\section*{Data Availability Statement}
All datasets used in this article have been published previously.

\section*{Conflict of Interest}
The authors have no conflicts to disclose.

\bibliography{literature}

\end{document}